\documentclass[aps,prb,reprint,superscriptaddress,amsmath,amssymb]{revtex4-2}
\usepackage{graphicx}
\usepackage{bm}
\usepackage{hyperref}

\usepackage{tikz}

\begin{document}
\title{Physical admissibility by construction in dispersive multipolar electromagnetic sheet models}
\author{Ravi S. Hegde}
\email{hegder@iitgn.ac.in}
\affiliation{13/322 B, Department of Electrical Engineering, Indian Institute of Technology Gandhinagar, Palaj, Gandhinagar 382355, Gujarat, India}
\begin{abstract}
Generalized sheet transition conditions (GSTCs) provide a powerful framework for modeling electromagnetic metasurfaces. Physical admissibility in these models is conventionally enforced through restrictions on the constitutive susceptibility tensors. However, the increasing complexity of multipolar and dispersive sheet models makes the formulation of such restrictions challenging. Here, we propose an alternative approach in which physical admissibility is enforced directly at the level of the effective surface admittance. We develop a general admittance formulation through quadrupolar order, accommodating spatial and temporal dispersion and unequal dielectric embeddings. Using sheet-level power and Lorentz reaction identities together with causal linear-response theory, we establish exact conditions for physical admissibility in the Laplace domain. We further derive an explicit construction of dispersive multipolar susceptibilities from prescribed admissible surface admittances throughout the open right half-plane at fixed tangential momentum. This construction guarantees constitutive closure regularity while retaining freedom in the allocation of the prescribed response among different multipolar moments, thereby separating physical admissibility from constitutive realization.
\end{abstract}
\maketitle
\section{Introduction}\label{sec:intro}

Electromagnetic metasurfaces are engineered interfaces that control the reflection, transmission, polarization, and wavefronts of electromagnetic waves. Their functionalities increasingly involve coupled control of polarization, spectral response, and angular scattering~\cite{Ataloglou2023,Zhao2022,Ghaneizadeh2024,Cuesta2025}. Modeling such interfaces without resolving their individual scattering elements requires effective descriptions capable of capturing the relevant electromagnetic interactions. Generalized sheet transition conditions (GSTCs) provide a framework for this purpose by relating field discontinuities to effective surface polarization sources~\cite{Kuester2003}. Extensions beyond the local dipolar approximation incorporate spatial dispersion through wavevector-dependent susceptibilities, higher multipolar moments, and field-gradient couplings~\cite{Rahmeier2023,Achouri2023Multipolar,Tiukuvaara2023,Monticone2025Roadmap}. These developments have enabled the description of structural symmetries, angle-invariant scattering, and the angular response of substrated metasurfaces~\cite{Achouri2023Symmetry,Yucel2025,Allahverdizadeh2025}.

An essential requirement of such effective models is that they respect the physical properties of the electromagnetic sheets they represent. For sheets intended to be passive and reciprocal, these properties include causality, passivity, and reciprocity. The transition conditions follow from Maxwell's equations, but the constitutive relations specifying the surface sources must additionally ensure these properties. In susceptibility-based GSTC formulations, physical admissibility is conventionally imposed through restrictions on the constitutive tensors, as exemplified by the reciprocity and losslessness relations derived for quadrupolar media~\cite{Achouri2021}. Establishing appropriate restrictions becomes increasingly involved as the constitutive description is extended to include higher multipolar moments, dielectric embeddings, and dispersion.

The electromagnetic embedding is an important consideration in formulating these restrictions. Generalizations of the dipolar GSTCs to dielectric interfaces have emphasized the role of the surrounding media in defining the effective constitutive parameters~\cite{Albooyeh2016}. More recent work has examined the prescription of the exciting fields and the associated hybridization of susceptibilities~\cite{Hansen2026}, as well as the treatment of normal polarizations at dielectric interfaces~\cite{HansenMiskovic2026}. These considerations are particularly relevant to multipolar descriptions, in which the moments are excited by both the fields and their spatial derivatives. At an interface between unequal media, these excitations can depend on the field discontinuities generated by the moments themselves, introducing constitutive feedback. Admissibility restrictions on the susceptibilities must therefore account for the excitation prescription and the electromagnetic embedding, as well as for the regularity of the resulting closure.

A related consideration is the nonuniqueness of the constitutive description. Equivalent realizations involving tangential and normal dipolar polarizations~\cite{Albooyeh2017}, and the representation dependence of effective parameters in bulk spatially dispersive media~\cite{Yaghjian2013,Skaar2019,Dirdal2015}, illustrate how different polarization descriptions can represent the same electromagnetic response. Similar questions arise in the reduction of cascaded local sheets to a single nonlocal sheet~\cite{Phaneuf2025} and in the interpretation of strongly spatially dispersive magnetoelectric coupling~\cite{Phaneuf2026}. In a multipolar sheet, the equivalent surface currents are determined by the combined contributions of the retained moments, and distinct susceptibility assignments can therefore produce the same current response. Restrictions imposed directly on individual susceptibility tensors must be formulated with care so as to preserve this freedom while ensuring the admissibility of the collective response.

These considerations acquire further significance when the model is intended to describe broadband behavior. Extended GSTCs allow the susceptibilities to depend on the tangential wavevector, for example through rational functions of the spatial frequency~\cite{Rahmeier2023}. When temporal dispersion is also present, the susceptibilities must describe a causal response across frequencies, and monochromatic power balance alone does not establish temporal passivity. The importance of stability in GSTC formulations has been recognized in Ref.~\cite{Lebbe2022}, while causality constraints on metasurface absorption have been investigated using surface conductivities~\cite{Valagiannopoulos2025}. More generally, the theory of passive linear response provides a well-established characterization of causal passive operators in terms of positive-real and related Herglotz functions~\cite{Welters2014,Bernland2011,Nedic2021,Srivastava2021}. For multipolar sheets, applying this theory requires a description of the collective response, including the constitutive feedback and the combined contributions of the moments.

Taken together, these developments motivate an alternative to imposing physical admissibility directly on the susceptibility tensors: one may first impose the physical constraints on the effective sheet response and subsequently determine a constitutive realization. Surface impedance and admittance formulations provide a natural basis for this approach. Tensorial descriptions are well established for bianisotropic dipolar sheets~\cite{BudhuGrbic2021,Szymanski2021,Ghaneizadeh2024}. Impedance formulations have also been employed in the synthesis of spatially dispersive metasurfaces~\cite{Asadchy2016}, including uniform nonlocal reflective surfaces~\cite{Zhuravlev2025Nonlocal}. The advantages of working directly with the response are illustrated by the synthesis of nonunique passive reciprocal tensor impedances for prescribed reflection transformations~\cite{Arnold2025}, and by the use of passive causal resonance models for broadband impedance synthesis~\cite{Fathnan2020Bandwidth}.

Extensions toward more general sheet models have proceeded along complementary directions. Surface impedance homogenization has incorporated selected dipolar and quadrupolar contributions in dielectric particle arrays~\cite{Monti2021MultipolarImpedance}, as well as power-preserving dipolar descriptions of disordered monolayers~\cite{Khezripour2026Impedance}. In the susceptibility-based GSTC literature, spatially dispersive models have been developed for synthesis and time-domain analysis, including joint spatial and temporal dispersion at dipolar order~\cite{Rahmeier2023,Dugan2024Synthesis,Rahmeier2024TimeDomain}. Causal convolution-based impedance GSTCs have also been developed for spatially local sheets~\cite{Kouroublakis2026Transient}. To the best of our knowledge, however, a general surface admittance formulation retaining the complete dipolar--quadrupolar constitutive closure, spatial and temporal dispersion, and unequal dielectric embeddings has not previously been combined with a complex-frequency admissibility characterization and an explicit inverse constitutive construction. Extending the response-based approach to this setting would bring the established methods of passive linear systems and network synthesis~\cite{Anderson1973,Kurokawa1965} to bear on a substantially more general class of multipolar sheet models.

In this work, building on the multipolar GSTCs of Refs.~\cite{Tiukuvaara2023,Achouri2023Multipolar}, we establish a general surface admittance formulation through quadrupolar order, accommodating spatial and temporal dispersion and unequal dielectric embeddings. The formulation retains all sixteen constitutive blocks, including normal moments, dipolar--quadrupolar cross-couplings, field-gradient excitations, and embedding-dependent feedback. Using sheet-level power and Lorentz reaction identities together with causal linear-response theory, we derive exact conditions for physical admissibility at both the admittance and susceptibility levels. The advantages of the response-based formulation are demonstrated by examining the limitations of conventional susceptibility restrictions in dielectric embeddings and the consequences of multipolar truncation for temporal passivity.

More importantly, we develop a constructive approach in which physically admissible surface admittances are prescribed and the corresponding dispersive multipolar susceptibilities are obtained explicitly. The construction guarantees constitutive closure regularity throughout the open right half-plane at fixed real tangential momentum, independently of the isotropic dielectric embedding. It also preserves the freedom to distribute the prescribed response among different multipolar moments, thereby separating physical admissibility from constitutive realization.

Section~\ref{sec:foundations} establishes the framework and Sec.~\ref{sec:conventional} develops its physical consequences and constructive realization. Supporting derivations are given in the appendices.

\section{Physical admissibility of multipolar sheet models}\label{sec:foundations}

We first reduce the general dipolar--quadrupolar constitutive closure to an effective surface admittance, then establish the physical constraints on this response and translate them into exact susceptibility conditions.

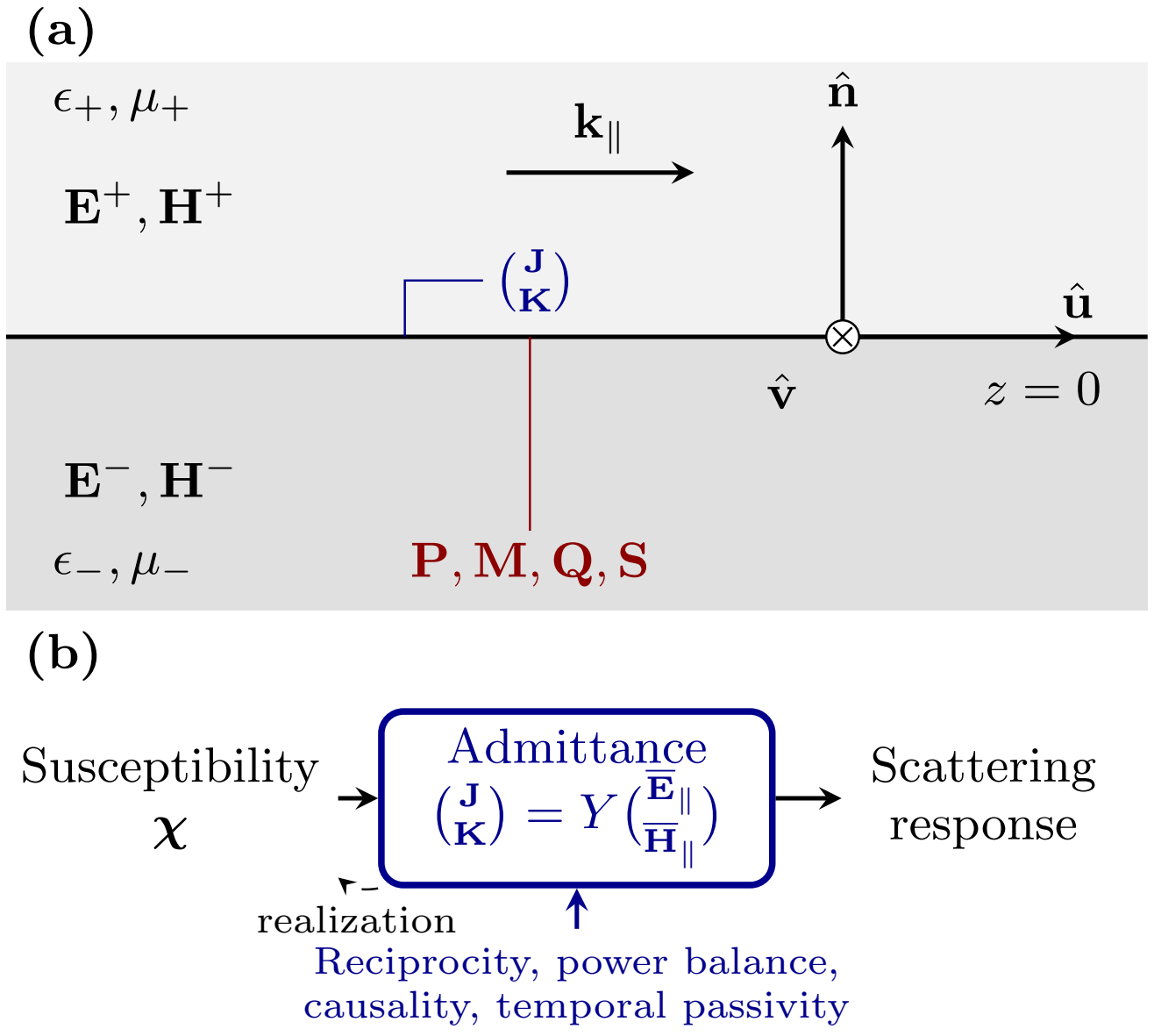
\begin{figure}[t]
\centering
\resizebox{\columnwidth}{!}{%
\begin{tikzpicture}[x=1cm,y=1cm,>=stealth,font=\small]
  \fill[black!5] (0,2.0) rectangle (7.3,3.75);
  \fill[black!12] (0,0.25) rectangle (7.3,2.0);
  \draw[thick] (0,2.0)--(7.3,2.0);
  \node[anchor=west] at (0.18,3.48) {$\epsilon_+,\mu_+$};
  \node[anchor=west] at (0.18,0.52) {$\epsilon_-,\mu_-$};
  \node[anchor=west] at (0.25,2.83) {$\mathbf E^+,\mathbf H^+$};
  \node[anchor=west] at (0.25,1.08) {$\mathbf E^-,\mathbf H^-$};
  \draw[->,thick] (5.35,2.0)--(5.35,3.35) node[above] {$\hat{\mathbf n}$};
  \draw[->,thick] (5.35,2.0)--(6.85,2.0) node[above] {$\hat{\mathbf u}$};
  \draw[fill=white] (5.35,2.0) circle (0.105);
  \draw (5.29,1.94)--(5.41,2.06);
  \draw (5.29,2.06)--(5.41,1.94);
  \node[anchor=north east] at (5.18,1.88) {$\hat{\mathbf v}$};
  \draw[->,thick] (3.20,3.05)--(4.40,3.05) node[midway,above] {$\mathbf k_\parallel$};
  \draw[thin,blue!55!black] (2.55,2.0)--(2.55,2.36)--(3.05,2.36);
  \node[anchor=west,inner sep=1pt,text=blue!55!black] at (3.08,2.36) {$\binom{\mathbf J}{\mathbf K}$};
  \draw[thin,red!55!black] (3.35,2.0)--(3.35,0.76);
  \node[anchor=north,inner sep=2pt,text=red!55!black] at (3.35,0.76) {$\mathbf P,\mathbf M,\mathbf Q,\mathbf S$};
  \node[anchor=east] at (7.12,1.67) {$z=0$};
  \node[anchor=west,font=\bfseries\small] at (0,3.95) {(a)};
  \node[anchor=west,font=\bfseries\small] at (0,-0.04) {(b)};
  \node[align=center,minimum width=1.8cm,minimum height=1.0cm] (chi) at (1.05,-0.95) {Susceptibility\\$\boldsymbol{\chi}$};
  \node[draw,rounded corners,very thick,blue!55!black,align=center,minimum width=2.5cm,minimum height=1.0cm] (adm) at (3.65,-0.95) {Admittance\\$\binom{\mathbf J}{\mathbf K}=Y\binom{\overline{\mathbf E}_\parallel}{\overline{\mathbf H}_\parallel}$};
  \node[align=center,minimum width=1.8cm,minimum height=1.0cm] (scat) at (6.25,-0.95) {Scattering\\response};
  \draw[->,thick] (chi.east)--(adm.west);
  \draw[->,thick] (adm.east)--(scat.west);
  \node[align=center,font=\scriptsize,text=blue!55!black] (laws) at (3.65,-2.15) {Reciprocity, power balance,\\causality, temporal passivity};
  \draw[->,blue!55!black,thick] (laws.north)--(adm.south);
  \draw[->,dashed] (adm.south west) to[bend left=25] node[below,font=\scriptsize] {realization} (chi.south east);
\end{tikzpicture}%
}
\caption{(a) Planar dipole--quadrupole sheet separating two dielectric half-spaces, with field traces, moments, and equivalent currents. (b) Constitutive representation, effective sheet admittance, and scattering response. For a specified embedding, a regular constitutive closure determines $Y$, which in turn determines scattering wherever the scattering problem is well posed. Physical constraints are imposed at the admittance level; the dashed arrow denotes the inverse constitutive realization problem, when solvable. The admittance acts on the average tangential fields of panel (a) and returns the same equivalent currents $\mathbf J,\mathbf K$ shown there.}
\label{fig:sheet_geometry}\label{fig:effective_admittance}
\end{figure}

Consider a planar sheet at $z=0$, with unit normal $\hat{\mathbf n}$ directed from the lower ($-$) to the upper ($+$) half-space. The time convention is $\exp(i\omega t)$. We write $\Delta\mathbf A=\mathbf A^+-\mathbf A^-$ and $\overline{\mathbf A}=(\mathbf A^++\mathbf A^-)/2$ for the jump and arithmetic mean of the field traces (Fig.~\ref{fig:sheet_geometry}).

\subsection{Generalized multipolar sheet admittance}

For a uniform sheet with tangential dependence $\exp(i\mathbf k_\parallel\cdot\mathbf r_\parallel)$, collect the dipole and quadrupole components into $m=(\mathbf P,\mathbf M,\operatorname{vec}\mathbf Q,\operatorname{vec}\mathbf S)^T\in\mathbb C^{24}$ before imposing tensor symmetries. Define $j=(\mathbf J,\mathbf K)^T\in\mathbb C^4$, where
\begin{equation}
 \mathbf J=\hat{\mathbf n}\times\Delta\mathbf H,\quad
 \mathbf K=-\hat{\mathbf n}\times\Delta\mathbf E .
 \label{eq:equivalent_currents}
\end{equation}
Define the average tangential field vector $x=(\overline{\mathbf E}_\parallel,\overline{\mathbf H}_\parallel)^T\in\mathbb C^4$. For a linear sheet invariant under tangential translations, we seek a regular response of the form
\begin{equation}
 j=Y(\omega,\mathbf k_\parallel)x,
 \label{eq:sheet_response_object}
\end{equation}
where $Y$ is the effective surface admittance. This current--field representation follows the established bianisotropic sheet impedance/admittance formalism \cite{BudhuGrbic2021}; surface impedance models incorporating dipolar and quadrupolar contributions have also been developed for dielectric metasurfaces \cite{Monti2021MultipolarImpedance}. Closures inducing the same $Y$ in a given embedding impose identical tangential transition relations. The tangential variables determine the normal fields and first derivatives through bulk Maxwell equations (Appendix~\ref{app:excitation}); this reduction does not discard normal moments.

With $\mathbf k=\mathbf k_\parallel$, $R\mathbf v=\hat{\mathbf n}\times\mathbf v$, and tensor blocks (normal rows written as column vectors)
\begin{align}
 \mathbf a_T&=\mathbf k T_{nn}-T_{\parallel\parallel}\mathbf k,\\
 b_T&=\mathbf k^T(T_{n\parallel}+T_{\parallel n}).
\end{align}
the dipole--quadrupole GSTCs read \cite{Tiukuvaara2023,Achouri2023Multipolar}
\begin{align}
 \hat{\mathbf n}\times\Delta\mathbf H
 &=i\omega\mathbf P_\parallel-iM_nR\mathbf k
 -\frac{\omega}{2}\mathbf a_Q\nonumber\\
 &\quad-\frac12 b_S R\mathbf k
 +\frac{\omega^2\epsilon_0\mu_0}{2}R S_{\parallel n},
 \label{eq:gstc_electric}\\
 -\hat{\mathbf n}\times\Delta\mathbf E
 &=i\omega\mu_0\mathbf M_\parallel
 +\frac{iP_n}{\epsilon_0}R\mathbf k
 -\frac{\omega\mu_0}{2}\mathbf a_S\nonumber\\
 &\quad+\frac{b_Q}{2\epsilon_0}R\mathbf k
 -\frac{\omega^2\mu_0}{2}R Q_{\parallel n}.
 \label{eq:gstc_magnetic}
\end{align}
Here $\epsilon_0,\mu_0$ are the source reference constants. These equations define the linear map from multipolar moments to equivalent surface currents,
\begin{equation}
 j=L(\omega,\mathbf k)m,\qquad
 L:\mathbb C^{24}\to\mathbb C^4 .
 \label{eq:moment_current_map}
\end{equation}
For source-free regular fields and $\omega\ne0$, the normal flux jumps are $\hat{\mathbf n}\cdot\Delta\mathbf D=-\mathbf k^T\mathbf J/\omega$ and $\hat{\mathbf n}\cdot\Delta\mathbf B=-\mathbf k^T\mathbf K/\omega$.

\paragraph{Constitutive model.}\label{sec:normalized_closure}
In the normalized coordinates of Eq.~(8) of Tiukuvaara \textit{et al.}~\cite{Tiukuvaara2023}, the full sixteen-block closure reads
\begin{equation}
\begin{gathered}
\begin{pmatrix}
 P_i/\epsilon_0\\ \eta_0 M_i\\ k_0Q_{il}/\epsilon_0\\ k_0\eta_0 S_{il}
\end{pmatrix}
=X
\begin{pmatrix}
 E_{{\rm av},j}\\ \eta_0 H_{{\rm av},j}\\
 \overline{\partial_k E_j}/(2k_0)\\ \eta_0\overline{\partial_k H_j}/(2k_0)
\end{pmatrix},\\[4pt]
X=\begin{pmatrix}
 \chi_{\rm ee}^{ij}&\chi_{\rm em}^{ij}&\chi_{\rm ee}^{\prime ijk}&\chi_{\rm em}^{\prime ijk}\\
 \chi_{\rm me}^{ij}&\chi_{\rm mm}^{ij}&\chi_{\rm me}^{\prime ijk}&\chi_{\rm mm}^{\prime ijk}\\
 Q_{\rm ee}^{ilj}&Q_{\rm em}^{ilj}&Q_{\rm ee}^{\prime iljk}&Q_{\rm em}^{\prime iljk}\\
 S_{\rm me}^{ilj}&S_{\rm mm}^{ilj}&S_{\rm me}^{\prime iljk}&S_{\rm mm}^{\prime iljk}
\end{pmatrix}.
\end{gathered}
\label{eq:published_closure}
\end{equation}
Here $k_0=\omega\sqrt{\epsilon_0\mu_0}$ and $\eta_0=\sqrt{\mu_0/\epsilon_0}$; repeated Cartesian excitation indices are summed. The subscript $\mathrm{av}$ denotes the acting-field average, with flux weighting only for normal field components as specified in Appendix~\ref{app:normalization}; overbars on field gradients denote averages of the physical one-sided derivatives. Appendix~\ref{app:excitation} specifies the flux-weighted normal acting values, physical one-sided gradients, normalized source map $F$, excitation map $G_{a,b}$ ($a=\epsilon/\epsilon_0$, $b=\mu/\mu_0$), and feedback $B_{\rm n}=D_{\rm j}H_{\rm j}$. The normalized admittance is Eq.~\eqref{eq:dielectric_admittance}. We use unnormalized physical variables for Laplace-domain statements to avoid frequency-dependent coordinate singularities.

\paragraph{Exact reduction.}
For a linear, time-invariant, tangentially uniform sheet, write $m=\boldsymbol{\chi} f$, where $f$ collects the unnormalized exciting fields and their gradients, and $\boldsymbol{\chi}$ is the susceptibility in physical coordinates, related to $X$ by Eq.~\eqref{eq:normalized_susceptibility}. The one-sided Maxwell eliminations and the excitation prescription give $f=Ax+Bj$ (Appendix~\ref{app:excitation}). In particular, the difference between the two half-space excitation maps generates the feedback term $Bj$; setting it to zero without justification changes the constitutive model. Substitution into $j=Lm$ yields
\begin{equation}
 (I_4-L\;\boldsymbol{\chi} B)j=L\;\boldsymbol{\chi} Ax .
 \label{eq:constitutive_feedback}
\end{equation}
We restrict the main analysis to regular closures, for which $I_4-L\;\boldsymbol{\chi} B$ is invertible. The unique current response to an arbitrary $x$ therefore realizes Eq.~\eqref{eq:sheet_response_object} with
\begin{equation}
 Y=(I_4-L\;\boldsymbol{\chi} B)^{-1}L\;\boldsymbol{\chi} A .
 \label{eq:effective_admittance}
\end{equation}
Equation~\eqref{eq:effective_admittance} is the general constitutive-to-admittance reduction central to this work. All operators may depend jointly on frequency and tangential momentum; all sixteen constitutive blocks, normal moments, gradient couplings and embedding-dependent feedback are retained. The inverse factor is absent in the conventional tangential dipolar model with $B=0$, for which $Y=L\boldsymbol{\chi}A$. The reduction is exact within the adopted multipolar model wherever the closure is regular. We now impose the physical constraints on this effective response.

\subsection{Physical admissibility of the effective response}\label{sec:admissibility}

Poynting's theorem and Lorentz reciprocity~\cite{Harrington2001,Achouri2021} provide the power and reaction balances needed to constrain the effective admittance. With $\Phi^\pm=\tfrac12\operatorname{Re}[\hat{\mathbf n}\cdot(\mathbf E^\pm\times\mathbf H^{\pm *})]$, the midpoint--jump expansion gives
\begin{align}
 p_{\rm s}&=\Phi^--\Phi^+\nonumber\\
 &=\tfrac12\operatorname{Re}\bigl(
 \overline{\mathbf E}_\parallel\cdot\mathbf J^*
 +\mathbf K\cdot\overline{\mathbf H}_\parallel^*\bigr)\nonumber\\
 &=\tfrac12\operatorname{Re}(x^\dagger Lm).
 \label{eq:surface_power}
\end{align}
For reciprocity, consider two independent states $a,b$ at the same frequency and define the normal Lorentz reaction flux
\begin{equation}
 \Psi_{ab}^{\pm}=\hat{\mathbf n}\cdot
 (\mathbf E_a^\pm\times\mathbf H_b^\pm
 -\mathbf E_b^\pm\times\mathbf H_a^\pm).
\end{equation}
The analogous bilinear midpoint--jump expansion gives
\begin{align}
 \mathcal R_{ab}&=\Psi_{ab}^--\Psi_{ab}^+\nonumber\\
 &=x_a^T\Sigma L_bm_b-x_b^T\Sigma L_am_a,
 \label{eq:surface_reaction}
\end{align}
where $\Sigma=\operatorname{diag}(I_2,-I_2)$ and $L_a=L(\omega,\mathbf k_a)$. Both balances depend on the multipolar moments only through their equivalent currents $j=Lm=Yx$ (Appendix~\ref{app:flux_pairings}). For a tangentially uniform sheet, spatial averaging selects equal momenta for the Hermitian power pairing and opposite momenta for the bilinear reaction pairing.

Consequently, the familiar admittance constraints of reciprocal and passive sheet theory~\cite{BudhuGrbic2021,Szymanski2021} apply to the effective multipolar response. Substituting $j=Yx$ into Eq.~\eqref{eq:surface_reaction}, reciprocity at real nonzero $\omega$ and real $\mathbf k$ is equivalent to
\begin{equation}
 \Sigma Y(\omega,-\mathbf k)=Y(\omega,\mathbf k)^T\Sigma .
 \label{eq:admissibility_reciprocity}
\end{equation}
Similarly, Eq.~\eqref{eq:surface_power} gives the condition for nonnegative absorbed power for every $x$,
\begin{equation}
 \operatorname{Herm}Y(\omega,\mathbf k)\succeq0 ,
 \label{eq:admissibility_power}
\end{equation}
where $\operatorname{Herm}Y=(Y+Y^\dagger)/2$. Monochromatic losslessness corresponds to equality. These conditions hold on the full independent average-field space, not merely for propagating excitations.

For a temporally dispersive sheet, admissibility must also be assessed at complex frequencies $s=\sigma+i\omega$, where $\omega$ is the real angular frequency. Let $\mathcal Y(s,\mathbf k)$ be the Laplace-domain admittance of the zero-state convolution response. Causal linear-response theory~\cite{Welters2014,Srivastava2021} requires a causal kernel and analyticity wherever its Laplace transform exists; here we require the transform throughout $\operatorname{Re}s>0$. Its regular boundary values are $Y(\omega,\mathbf k)=\mathcal Y(0^++i\omega,\mathbf k)$ under our time convention. A real temporal-spatial kernel requires $\mathcal Y(\bar s,-\mathbf k)=\overline{\mathcal Y(s,\mathbf k)}$, while reciprocity extends to the analytic response as $\Sigma\mathcal Y(s,-\mathbf k)=\mathcal Y(s,\mathbf k)^T\Sigma$. Standard dispersion relations follow under the appropriate decay and boundary regularity assumptions~\cite{Welters2014}.

\begin{figure}[t]
 \centering
 \resizebox{\columnwidth}{!}{%
 \begin{tikzpicture}[x=1cm,y=1cm,>=stealth,font=\small]
  \fill[blue!9] (0,-1.42) rectangle (3.65,1.42);
  \draw[->] (-3.25,0) -- (3.83,0) node[below left] {$\sigma=\operatorname{Re}s$};
  \draw[blue!65!black,thick,->] (0,-1.55) -- (0,1.72) node[above,black] {$\omega=\operatorname{Im}s$};
  \node[align=center,text=blue!65!black] at (-1.67,0.89) {Monochromatic\\response};
  \node[text=blue!65!black] at (-1.67,0.31) {$s=i\omega$};
  \draw[blue!65!black,->] (-0.72,0.57) -- (-0.06,0.57);
  \node[align=center,fill=blue!9,inner sep=4pt] at (1.82,0)
   {Temporal passivity\\[3pt]
    $\mathcal Y$ analytic\\[2pt]
    $\operatorname{Herm}\mathcal Y\succeq0$\\[3pt]
    throughout $\sigma>0$};
  \draw[->,thin] (-0.79,-0.76) -- (-0.065,-0.065);
  \node[align=center] at (-1.74,-0.83) {Static point\\$s=0$};
  \fill[black] (0,0) circle (1.6pt);
 \end{tikzpicture}%
 }
 \caption{Domains of physical admissibility conditions in the complex-frequency plane. At fixed real tangential momentum $\mathbf k$, each point $s$ represents an evaluation of the $4\times4$ admittance matrix $\mathcal Y(s,\mathbf k)$ wherever defined. Temporal passivity requires analyticity and a positive semidefinite Hermitian part throughout the open right half-plane; regular imaginary-axis boundary values yield the monochromatic response. The origin may require separate static analysis.}
 \label{fig:laplace_admissibility}
\end{figure}
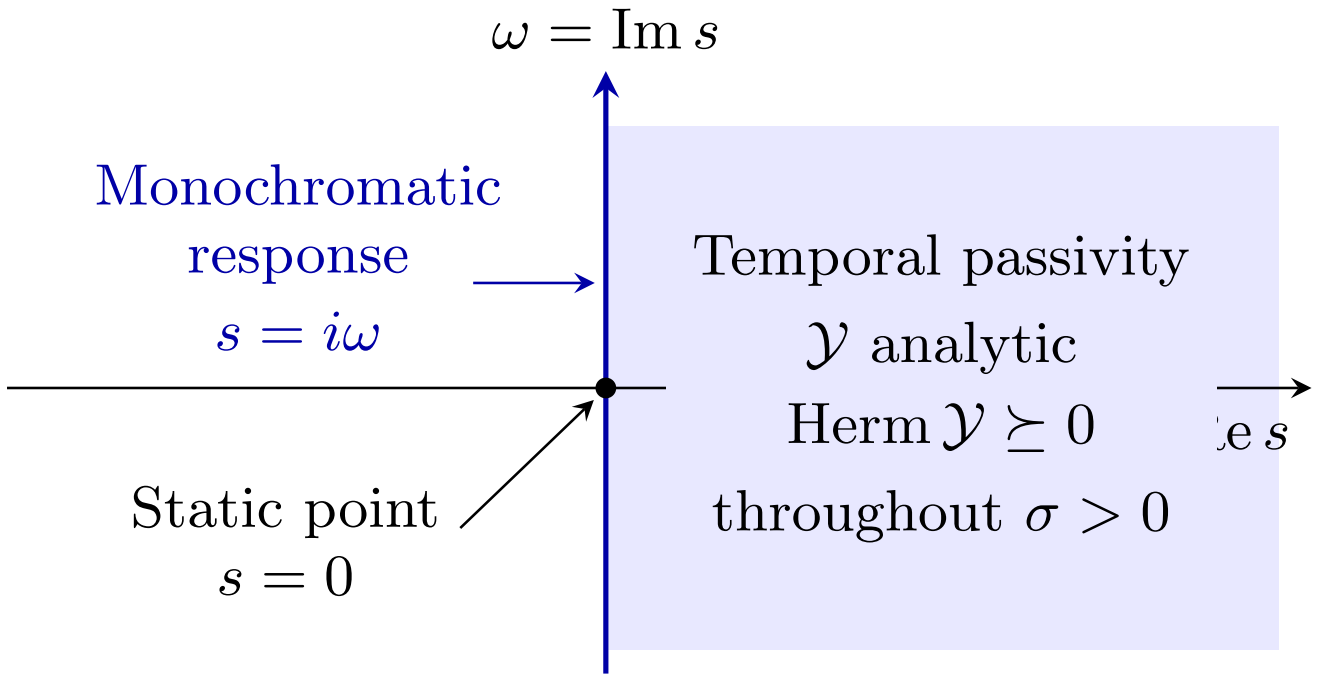

For causal convolution operators on smooth compactly supported test fields, temporal passivity (nonnegative work on the initially unexcited sheet up to every time) is characterized by positive-realness, as in passive network and linear-response theory~\cite{Anderson1973,Welters2014,Bernland2011,Nedic2021}:
\begin{equation}
 \operatorname{Herm}\mathcal Y(s,\mathbf k)\succeq0,
 \qquad \operatorname{Re}s>0 .
 \label{eq:positive_real}
\end{equation}
For a spatially dispersive sheet this condition applies almost everywhere in $\mathbf k$, or at every momentum when the response is continuous in $\mathbf k$. Unlike Eq.~\eqref{eq:admissibility_power}, Eq.~\eqref{eq:positive_real} tests temporal passivity throughout the causal half-plane (Fig.~\ref{fig:laplace_admissibility}). Nonnegative absorbed power at regular real frequencies alone is insufficient, as the truncation example in Sec.~\ref{sec:multipolar_extension} demonstrates.

The effective admittance thus provides a common representation in which to assess both the monochromatic and temporal properties of a sheet, irrespective of the particular constitutive assignment that generates it. The remaining question is how these physical requirements constrain the underlying multipolar susceptibilities.

\subsection{Exact admissibility conditions on multipolar susceptibilities}\label{sec:constitutive_admissibility}

The conditions of Sec.~\ref{sec:admissibility} are expressed in terms of the effective admittance. To apply them directly to the susceptibility tensors, we return to the current equation~\eqref{eq:constitutive_feedback}. Write it as $Uj=Vx$, with
\begin{equation}
 U=\underbrace{I_4-L\boldsymbol{\chi}B}_{\text{feedback matrix}},
 \qquad
 V=\underbrace{L\boldsymbol{\chi}A}_{\text{direct current map}} .
 \label{eq:closure_UV}
\end{equation}
The $4\times4$ matrix $V$ describes the currents induced directly by the average tangential fields, while $U$ accounts for the additional excitation due to the field discontinuities. In the absence of feedback ($B=0$), $U=I_4$ and the admittance is simply $V$. In general, a regular closure gives $Y=U^{-1}V$.

Substituting this admittance into the power condition~\eqref{eq:admissibility_power} and using congruence by the invertible matrix $U$ gives the equivalent condition
\begin{equation}
 VU^\dagger+UV^\dagger\succeq0 ,
 \label{eq:chi_power}
\end{equation}
with equality for losslessness. Similarly, substitution into the reciprocity condition~\eqref{eq:admissibility_reciprocity} and multiplication by the feedback matrices gives
\begin{equation}
 V_-\Sigma U_+^T=U_-\Sigma V_+^T ,
 \label{eq:chi_reciprocity}
\end{equation}
where $U_\pm=U(\omega,\pm\mathbf k)$ and $V_\pm=V(\omega,\pm\mathbf k)$ denote evaluation at opposite tangential momenta, not the two half-spaces. Appendix~\ref{app:constitutive_tests} gives the algebraic details of both equivalences.

For specified $A,B,L$, these exact tests constrain $\boldsymbol{\chi}$ without assuming intrinsic tensor symmetries. Feedback introduces quadratic susceptibility terms; when $B=0$, the conditions are linear.

For temporally dispersive sheets, the same algebra applies in the Laplace domain ($\operatorname{Re}s>0$). Write $\mathcal L,\mathcal A,\mathcal B$ for the Laplace-domain source and excitation maps and $\mathcal U,\mathcal V$ for the corresponding feedback and direct current matrices. For a regular closure, temporal passivity is tested by Eq.~\eqref{eq:chi_power} with $U,V$ replaced by $\mathcal U,\mathcal V$ throughout $\operatorname{Re}s>0$; reciprocity is tested by Eq.~\eqref{eq:chi_reciprocity} with the same replacement. The induced admittance $\mathcal Y=\mathcal U^{-1}\mathcal V$ must also satisfy the analyticity, reality and causal-kernel requirements of Sec.~\ref{sec:admissibility}.

For analytic $\mathcal U,\mathcal V$, invertibility of $\mathcal U$ throughout the right half-plane ensures response analyticity, but does not exclude imaginary-axis poles or instantaneous derivative terms. For a proper rational admittance at fixed momentum, strictly left-half-plane poles characterize bounded-input bounded-output stability and, for minimal realizations, asymptotic internal stability~\cite{Anderson1973}. Nonminimal realizations may conceal internal modes through cancellations at zeros of $\det\mathcal U$; stability of the coupled exterior Maxwell problem remains separate.

The resulting framework distinguishes physical constraints on the response from regularity and admissibility conditions on its constitutive realization.

\section{Physical implications for dispersive multipolar sheets}\label{sec:conventional}

We now examine three consequences of the response-based framework of Sec.~\ref{sec:foundations}: embedding-dependent susceptibility conditions, loss of temporal passivity under multipolar truncation, and the constructive realization of physically admissible responses.

\subsection{Embedding dependence of reciprocity and losslessness conditions}\label{sec:dielectric_failure}

The reciprocity and losslessness restrictions on multipolar susceptibilities derived in Ref.~\cite{Achouri2021} are commonly expressed as intrinsic tensor symmetries; the reciprocity restrictions are also employed for substrated metasurfaces~\cite{Tiukuvaara2023}. Are these restrictions sufficient in a dielectric embedding? The question concerns the conditions imposed on the susceptibilities, irrespective of how their values are obtained.

\paragraph{Vacuum and dielectric embeddings.}
For identical isotropic half-spaces, the constitutive feedback vanishes and Eq.~\eqref{eq:dielectric_admittance} reduces to $\tilde Y=FXG_{a,b}$. In vacuum, with $G=G_{1,1}$, the source and excitation maps satisfy (Appendix~\ref{app:applications})
\begin{equation}
 F=ik_0G^\dagger .
 \label{eq:vacuum_source_adjoint}
\end{equation}
The admittance therefore takes the form
\begin{equation}
 \tilde Y=ik_0G^\dagger XG .
 \label{eq:vacuum_admittance}
\end{equation}
The following susceptibility restrictions are sufficient for losslessness and reciprocity, respectively:
\begin{align}
 X(\mathbf k)&=X(\mathbf k)^\dagger,\\
 X(-\mathbf k)&=DX(\mathbf k)^TD,
 \label{eq:vacuum_sufficient}
\end{align}
where $D=\operatorname{diag}(-I_3,I_3,-I_9,I_9)$. For momentum-independent susceptibilities with the corresponding intrinsic index symmetries, these recover Eqs.~(38)--(40) and (27)--(29) of Ref.~\cite{Achouri2021}. They are sufficient but not necessary because the physical conditions constrain the compressed response.

In a dielectric, the field gradients exciting the quadrupoles depend on the material parameters, e.g., $\partial_z H_y=-i\omega\epsilon E_x$ at normal incidence. The source map retains its vacuum reference constants under the adopted GSTCs, whereas the excitation map changes. Consequently, $\tilde Y=ik_0G^\dagger XG_{a,b}$ no longer has the vacuum adjoint structure. The exact losslessness and reciprocity conditions now involve the embedding parameters explicitly [Eq.~\eqref{eq:dielectric_exact_conditions}].

For unequal half-spaces, including a vacuum--substrate interface, the difference between the one-sided excitation maps introduces additional feedback through the field jumps. The admittance is then given by Eq.~\eqref{eq:dielectric_admittance}, with the embedding-dependent average excitation $C$ and feedback $B_{\rm n}$. The corresponding susceptibility conditions are those of Sec.~\ref{sec:constitutive_admissibility}, rather than the vacuum tensor restrictions.

\paragraph{Counterexample.}
Consider the following susceptibility at normal incidence in lossless, nonmagnetic half-spaces:
\begin{equation}
 X=\lambda vv^\dagger,\qquad
 v=e_{P_x}+i(e_{S_{xz}}+e_{S_{zx}}),
 \label{eq:dielectric_counterexample_assignment}
\end{equation}
where $\lambda$ is real and nonzero and the $e$'s are the normalized moment coordinate vectors. This electric dipole--magnetic quadrupole coupling satisfies both vacuum restrictions, since $X=X^\dagger$ and $D\overline v=-v$. The quadrupole is symmetric and traceless. The electric admittance block is (Appendix~\ref{app:unequal_dielectric_example})
\begin{equation}
 \tilde Y_{\rm ee}=ik_0\lambda
 \begin{pmatrix}1&\bar a/2\\1/2&\bar a/4\end{pmatrix},
 \qquad \bar a=\frac{a_++a_-}{2},
\end{equation}
where $a_\pm=\epsilon_\pm/\epsilon_0$. For this assignment, the closure is regular and feedback does not alter the resulting current; the expression holds for identical as well as unequal media.

When $\bar a\ne1$, the block is nonsymmetric and its Hermitian part has eigenvalues $\pm|k_0\lambda(\bar a-1)|/4$. Thus a susceptibility satisfying the conventional restrictions can violate reciprocity and even predict negative absorbed power in a lossless dielectric embedding. Figure~\ref{fig:admissibility_examples}(a) illustrates this failure for both identical media and vacuum--substrate interfaces.

The counterexample establishes that the conventional tensor restrictions do not suffice to enforce reciprocity and losslessness in dielectric embeddings, even if the susceptibilities are assigned specifically for that environment. The admissibility conditions must instead be derived using the excitation and source maps of the intended embedding.

\subsection{Failure of temporal passivity under multipolar truncation}\label{sec:multipolar_extension}\label{sec:observable_equivalence}

Even in a fixed embedding, physical admissibility may depend on which moments are retained in the constitutive model. Higher multipolar moments are often introduced to improve accuracy; can they also be essential to temporal passivity? In a dispersive model, individual moments contribute collectively to the equivalent sheet currents. Truncating the constitutive representation can therefore remove cancellations required for temporal passivity, even if the resulting model still predicts nonnegative absorbed power at every regular real frequency.

\paragraph{Reduction versus truncation.}
An exact reduction to a lower-order representation preserves the admittance and hence its physical properties. For example, in vacuum the normal dipolar contributions can be absorbed into momentum-dependent tangential coefficients, recovering the spatially dispersive dipolar description of Phaneuf and Mojabi~\cite{Phaneuf2026} in the TE sector (Appendix~\ref{app:te_reduction}). Simply deleting higher moments while retaining the remaining susceptibilities is a different operation: it changes the admittance. The distinction becomes particularly important for temporal passivity, which constrains the entire causal half-plane ($\operatorname{Re}s>0$) rather than only the regular real-frequency axis.

\paragraph{Counterexample.}
In the Laplace domain, consider a TE-preserving sheet in vacuum with $\mathbf k=k\hat{\mathbf x}$, $k\ne0$, and nonzero moments
\begin{align}
 M_z&=k^2 a(s)\overline H_z,\\
 S_{xz}=S_{zx}&=-ik\,a(s)\overline H_z,
 \label{eq:te_passivity_restoration}
\end{align}
where $a(s)=F_0/(s^2+\gamma s+\omega_0^2)$, with $F_0,\gamma,\omega_0>0$. The two moments share the same causal Lorentz kernel. Their contributions to the electric sheet current are, from the multipolar GSTCs,
\begin{equation}
 J_y=-ikM_z-\left(k^2+\frac{s^2}{2c^2}\right)S_{xz},
 \qquad c^{-2}=\epsilon_0\mu_0 .
\end{equation}
The spatial term proportional to $k^2 S_{xz}$ cancels the normal magnetic dipole contribution, leaving a term proportional to $s^2 S_{xz}$. Using $\overline H_z=-ik\overline E_y/(s\mu_0)$, the admittances of the full sheet and of its dipolar truncation are
\begin{align}
 \mathcal Y_{\rm multi}&=\operatorname{diag}
 \left(\frac{k^2s a(s)}{2\mu_0c^2},0\right),\\
 \mathcal Y_{\rm dip}&=\operatorname{diag}
 \left(-\frac{k^4a(s)}{\mu_0s},0\right).
 \label{eq:te_truncation_admittances}
\end{align}
Both responses are reciprocal, satisfy the temporal reality condition and are analytic in the open right half-plane. On the regular real-frequency axis ($s=i\omega$, $\omega\ne0$), the dipolar admittance differs from the full admittance by the positive factor $2c^2k^2/\omega^2$. Both therefore satisfy the monochromatic power condition~\eqref{eq:admissibility_power}.

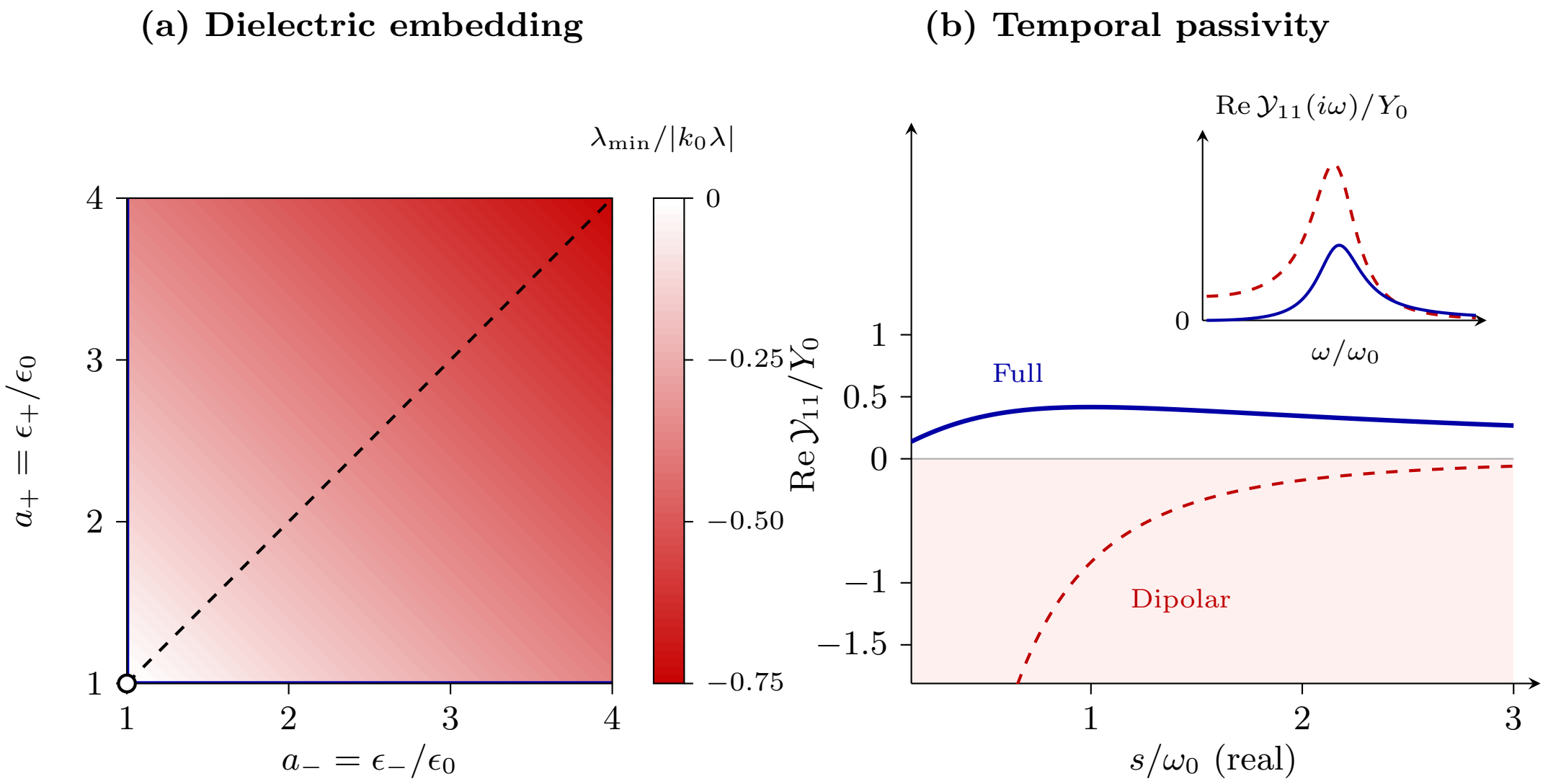
\begin{figure*}[t]
\centering
\begin{tikzpicture}[x=1.65cm,y=1.65cm,font=\small,>=stealth]
\begin{scope}
\node[anchor=west,font=\bfseries] at (0,3.65) {(a) Dielectric embedding};
\colorlet{embeddingtone}{red!78!black}
\fill[white] (0,0) rectangle (2.7,2.7);
\begin{scope}
 \clip (0,0) rectangle (2.7,2.7);
 \foreach \n in {1,...,60} {
  \pgfmathsetmacro{\threshold}{5.4*\n/60}
  \pgfmathsetmacro{\pct}{100*\n/60}
  \fill[embeddingtone!\pct!white]
   (\threshold,0)--(5.4,0)--(5.4,5.4)--(0,5.4)--(0,\threshold)--cycle;
 }
 \draw[black,dashed,thick] (0,0)--(2.7,2.7);
 \draw[blue!70!black,very thick] (0,2.7)--(0,0)--(2.7,0);
\end{scope}
\draw[black,thin] (0,0) rectangle (2.7,2.7);
\foreach \a in {1,2,3,4} {
 \draw ({.9*(\a-1)},0)--++(0,-.06) node[below] {$\a$};
 \draw (0,{.9*(\a-1)})--++(-.06,0) node[left] {$\a$};
}
\node at (1.35,-.43) {$a_-=\epsilon_-/\epsilon_0$};
\node[rotate=90] at (-.58,1.35) {$a_+=\epsilon_+/\epsilon_0$};
\filldraw[fill=white,draw=black,thick] (0,0) circle[radius=2.2pt];
\foreach \n in {0,...,59} {
 \pgfmathsetmacro{\pct}{100*\n/60}
 \fill[embeddingtone!\pct!white]
 (2.93,{2.7*(1-(\n+1)/60)}) rectangle (3.10,{2.7*(1-\n/60)});
}
\draw[black,thin] (2.93,0) rectangle (3.10,2.7);
\foreach \m/\lab in {0/0,1/{-0.25},2/{-0.50},3/{-0.75}} {
 \draw (3.10,{2.7-.9*\m})--++(.05,0) node[right,font=\scriptsize] {$\lab$};
}
\node[font=\scriptsize] at (2.98,3.03) {$\lambda_{\min}/|k_0\lambda|$};
\end{scope}
\begin{scope}[xshift=7.2cm]
\node[anchor=west,font=\bfseries] at (0,3.65) {(b) Temporal passivity};
\fill[red!6] (0,0) rectangle (3.35,1.25);
\draw[gray!65] (0,1.25)--(3.35,1.25);
\draw[->] (0,0)--(3.5,0);
\draw[->] (0,0)--(0,3.12);
\foreach \v in {-1.5,-1,0,0.5,1} {
 \draw (0,{1.25+.69*\v})--++(-.06,0) node[left] {$\v$};
}
\foreach \v in {1,2,3} {
 \draw ({(\v-.15)/2.85*3.35},0)--++(0,-.06) node[below] {$\v$};
}
\node at (1.68,-.43) {$s/\omega_0$ (real)};
\node[rotate=90] at (-.59,1.5) {$\operatorname{Re}\mathcal Y_{11}/Y_0$};
\begin{scope}
\clip (0,0) rectangle (3.35,3.05);
\draw[red!75!black,dashed,thick] (0.0000,-7.2488)--(0.0223,-6.1993)--(0.0447,-5.3598)--(0.0670,-4.6725)--(0.0893,-4.0994)--(0.1117,-3.6140)--(0.1340,-3.1976)--(0.1563,-2.8365)--(0.1787,-2.5202)--(0.2010,-2.2411)--(0.2233,-1.9929)--(0.2457,-1.7709)--(0.2680,-1.5711)--(0.2903,-1.3906)--(0.3127,-1.2266)--(0.3350,-1.0771)--(0.3573,-0.9404)--(0.3797,-0.8149)--(0.4020,-0.6994)--(0.4243,-0.5927)--(0.4467,-0.4941)--(0.4690,-0.4026)--(0.4913,-0.3176)--(0.5137,-0.2385)--(0.5360,-0.1647)--(0.5583,-0.0958)--(0.5807,-0.0314)--(0.6030,0.0290)--(0.6253,0.0857)--(0.6477,0.1389)--(0.6700,0.1890)--(0.6923,0.2361)--(0.7147,0.2805)--(0.7370,0.3223)--(0.7593,0.3619)--(0.7817,0.3992)--(0.8040,0.4345)--(0.8263,0.4680)--(0.8487,0.4997)--(0.8710,0.5297)--(0.8933,0.5582)--(0.9157,0.5853)--(0.9380,0.6109)--(0.9603,0.6354)--(0.9827,0.6586)--(1.0050,0.6807)--(1.0273,0.7018)--(1.0497,0.7218)--(1.0720,0.7409)--(1.0943,0.7592)--(1.1167,0.7766)--(1.1390,0.7932)--(1.1613,0.8091)--(1.1837,0.8242)--(1.2060,0.8387)--(1.2283,0.8526)--(1.2507,0.8659)--(1.2730,0.8786)--(1.2953,0.8908)--(1.3177,0.9024)--(1.3400,0.9136)--(1.3623,0.9243)--(1.3847,0.9346)--(1.4070,0.9445)--(1.4293,0.9539)--(1.4517,0.9630)--(1.4740,0.9718)--(1.4963,0.9802)--(1.5187,0.9882)--(1.5410,0.9960)--(1.5633,1.0035)--(1.5857,1.0107)--(1.6080,1.0176)--(1.6303,1.0242)--(1.6527,1.0307)--(1.6750,1.0368)--(1.6973,1.0428)--(1.7197,1.0486)--(1.7420,1.0541)--(1.7643,1.0594)--(1.7867,1.0646)--(1.8090,1.0696)--(1.8313,1.0744)--(1.8537,1.0790)--(1.8760,1.0835)--(1.8983,1.0878)--(1.9207,1.0920)--(1.9430,1.0961)--(1.9653,1.1000)--(1.9877,1.1038)--(2.0100,1.1074)--(2.0323,1.1110)--(2.0547,1.1144)--(2.0770,1.1177)--(2.0993,1.1209)--(2.1217,1.1240)--(2.1440,1.1271)--(2.1663,1.1300)--(2.1887,1.1328)--(2.2110,1.1356)--(2.2333,1.1382)--(2.2557,1.1408)--(2.2780,1.1433)--(2.3003,1.1457)--(2.3227,1.1481)--(2.3450,1.1504)--(2.3673,1.1526)--(2.3897,1.1548)--(2.4120,1.1569)--(2.4343,1.1589)--(2.4567,1.1609)--(2.4790,1.1628)--(2.5013,1.1647)--(2.5237,1.1665)--(2.5460,1.1683)--(2.5683,1.1700)--(2.5907,1.1717)--(2.6130,1.1733)--(2.6353,1.1749)--(2.6577,1.1764)--(2.6800,1.1779)--(2.7023,1.1794)--(2.7247,1.1808)--(2.7470,1.1822)--(2.7693,1.1835)--(2.7917,1.1848)--(2.8140,1.1861)--(2.8363,1.1873)--(2.8587,1.1886)--(2.8810,1.1897)--(2.9033,1.1909)--(2.9257,1.1920)--(2.9480,1.1931)--(2.9703,1.1942)--(2.9927,1.1952)--(3.0150,1.1963)--(3.0373,1.1972)--(3.0597,1.1982)--(3.0820,1.1992)--(3.1043,1.2001)--(3.1267,1.2010)--(3.1490,1.2019)--(3.1713,1.2027)--(3.1937,1.2036)--(3.2160,1.2044)--(3.2383,1.2052)--(3.2607,1.2060)--(3.2830,1.2067)--(3.3053,1.2075)--(3.3277,1.2082)--(3.3500,1.2089);
\draw[blue!65!black,very thick] (0.0000,1.3456)--(0.0223,1.3564)--(0.0447,1.3668)--(0.0670,1.3769)--(0.0893,1.3866)--(0.1117,1.3960)--(0.1340,1.4050)--(0.1563,1.4136)--(0.1787,1.4219)--(0.2010,1.4299)--(0.2233,1.4374)--(0.2457,1.4447)--(0.2680,1.4515)--(0.2903,1.4581)--(0.3127,1.4643)--(0.3350,1.4702)--(0.3573,1.4757)--(0.3797,1.4810)--(0.4020,1.4859)--(0.4243,1.4906)--(0.4467,1.4950)--(0.4690,1.4991)--(0.4913,1.5029)--(0.5137,1.5065)--(0.5360,1.5098)--(0.5583,1.5129)--(0.5807,1.5157)--(0.6030,1.5183)--(0.6253,1.5208)--(0.6477,1.5230)--(0.6700,1.5250)--(0.6923,1.5269)--(0.7147,1.5285)--(0.7370,1.5300)--(0.7593,1.5314)--(0.7817,1.5326)--(0.8040,1.5336)--(0.8263,1.5345)--(0.8487,1.5353)--(0.8710,1.5359)--(0.8933,1.5364)--(0.9157,1.5369)--(0.9380,1.5372)--(0.9603,1.5374)--(0.9827,1.5375)--(1.0050,1.5375)--(1.0273,1.5374)--(1.0497,1.5373)--(1.0720,1.5371)--(1.0943,1.5368)--(1.1167,1.5364)--(1.1390,1.5360)--(1.1613,1.5355)--(1.1837,1.5350)--(1.2060,1.5344)--(1.2283,1.5337)--(1.2507,1.5331)--(1.2730,1.5323)--(1.2953,1.5315)--(1.3177,1.5307)--(1.3400,1.5299)--(1.3623,1.5290)--(1.3847,1.5281)--(1.4070,1.5272)--(1.4293,1.5262)--(1.4517,1.5252)--(1.4740,1.5242)--(1.4963,1.5232)--(1.5187,1.5221)--(1.5410,1.5211)--(1.5633,1.5200)--(1.5857,1.5189)--(1.6080,1.5178)--(1.6303,1.5167)--(1.6527,1.5155)--(1.6750,1.5144)--(1.6973,1.5132)--(1.7197,1.5121)--(1.7420,1.5109)--(1.7643,1.5097)--(1.7867,1.5085)--(1.8090,1.5074)--(1.8313,1.5062)--(1.8537,1.5050)--(1.8760,1.5038)--(1.8983,1.5026)--(1.9207,1.5014)--(1.9430,1.5002)--(1.9653,1.4990)--(1.9877,1.4978)--(2.0100,1.4966)--(2.0323,1.4954)--(2.0547,1.4943)--(2.0770,1.4931)--(2.0993,1.4919)--(2.1217,1.4907)--(2.1440,1.4895)--(2.1663,1.4884)--(2.1887,1.4872)--(2.2110,1.4860)--(2.2333,1.4849)--(2.2557,1.4837)--(2.2780,1.4826)--(2.3003,1.4814)--(2.3227,1.4803)--(2.3450,1.4791)--(2.3673,1.4780)--(2.3897,1.4769)--(2.4120,1.4758)--(2.4343,1.4747)--(2.4567,1.4736)--(2.4790,1.4725)--(2.5013,1.4714)--(2.5237,1.4703)--(2.5460,1.4692)--(2.5683,1.4681)--(2.5907,1.4671)--(2.6130,1.4660)--(2.6353,1.4649)--(2.6577,1.4639)--(2.6800,1.4629)--(2.7023,1.4618)--(2.7247,1.4608)--(2.7470,1.4598)--(2.7693,1.4588)--(2.7917,1.4578)--(2.8140,1.4568)--(2.8363,1.4558)--(2.8587,1.4548)--(2.8810,1.4538)--(2.9033,1.4528)--(2.9257,1.4519)--(2.9480,1.4509)--(2.9703,1.4500)--(2.9927,1.4490)--(3.0150,1.4481)--(3.0373,1.4472)--(3.0597,1.4462)--(3.0820,1.4453)--(3.1043,1.4444)--(3.1267,1.4435)--(3.1490,1.4426)--(3.1713,1.4417)--(3.1937,1.4408)--(3.2160,1.4400)--(3.2383,1.4391)--(3.2607,1.4382)--(3.2830,1.4374)--(3.3053,1.4365)--(3.3277,1.4357)--(3.3500,1.4348);
\end{scope}
\node[anchor=west,text=red!75!black,font=\scriptsize] at (1.15,.46) {Dipolar};
\node[anchor=west,text=blue!65!black,font=\scriptsize] at (.38,1.73) {Full};
\begin{scope}[shift={(1.62,2.02)}]
\fill[white] (-.18,-.22) rectangle (1.68,1.2);
\draw[->,thin] (0,0)--(1.58,0);
\draw[->,thin] (0,0)--(0,1.06);
\draw[red!75!black,dashed,thick] (0.0228,0.1340)--(0.0328,0.1343)--(0.0428,0.1346)--(0.0527,0.1350)--(0.0627,0.1355)--(0.0727,0.1361)--(0.0827,0.1368)--(0.0927,0.1376)--(0.1027,0.1384)--(0.1126,0.1394)--(0.1226,0.1404)--(0.1326,0.1416)--(0.1426,0.1429)--(0.1526,0.1443)--(0.1625,0.1458)--(0.1725,0.1474)--(0.1825,0.1491)--(0.1925,0.1510)--(0.2025,0.1530)--(0.2124,0.1552)--(0.2224,0.1575)--(0.2324,0.1599)--(0.2424,0.1626)--(0.2524,0.1654)--(0.2624,0.1683)--(0.2723,0.1715)--(0.2823,0.1749)--(0.2923,0.1785)--(0.3023,0.1823)--(0.3123,0.1864)--(0.3222,0.1908)--(0.3322,0.1954)--(0.3422,0.2003)--(0.3522,0.2056)--(0.3622,0.2112)--(0.3721,0.2171)--(0.3821,0.2235)--(0.3921,0.2303)--(0.4021,0.2376)--(0.4121,0.2453)--(0.4221,0.2536)--(0.4320,0.2625)--(0.4420,0.2720)--(0.4520,0.2821)--(0.4620,0.2930)--(0.4720,0.3047)--(0.4819,0.3173)--(0.4919,0.3307)--(0.5019,0.3451)--(0.5119,0.3606)--(0.5219,0.3772)--(0.5318,0.3950)--(0.5418,0.4141)--(0.5518,0.4346)--(0.5618,0.4565)--(0.5718,0.4798)--(0.5818,0.5046)--(0.5917,0.5309)--(0.6017,0.5586)--(0.6117,0.5877)--(0.6217,0.6180)--(0.6317,0.6492)--(0.6416,0.6809)--(0.6516,0.7127)--(0.6616,0.7438)--(0.6716,0.7737)--(0.6816,0.8013)--(0.6915,0.8257)--(0.7015,0.8458)--(0.7115,0.8606)--(0.7215,0.8692)--(0.7315,0.8710)--(0.7415,0.8655)--(0.7514,0.8528)--(0.7614,0.8332)--(0.7714,0.8074)--(0.7814,0.7763)--(0.7914,0.7411)--(0.8013,0.7029)--(0.8113,0.6630)--(0.8213,0.6223)--(0.8313,0.5818)--(0.8413,0.5422)--(0.8513,0.5040)--(0.8612,0.4676)--(0.8712,0.4332)--(0.8812,0.4010)--(0.8912,0.3711)--(0.9012,0.3433)--(0.9111,0.3177)--(0.9211,0.2941)--(0.9311,0.2725)--(0.9411,0.2526)--(0.9511,0.2344)--(0.9610,0.2177)--(0.9710,0.2024)--(0.9810,0.1884)--(0.9910,0.1755)--(1.0010,0.1637)--(1.0110,0.1529)--(1.0209,0.1430)--(1.0309,0.1338)--(1.0409,0.1254)--(1.0509,0.1176)--(1.0609,0.1105)--(1.0708,0.1039)--(1.0808,0.0977)--(1.0908,0.0921)--(1.1008,0.0868)--(1.1108,0.0820)--(1.1207,0.0775)--(1.1307,0.0733)--(1.1407,0.0693)--(1.1507,0.0657)--(1.1607,0.0623)--(1.1707,0.0591)--(1.1806,0.0561)--(1.1906,0.0534)--(1.2006,0.0508)--(1.2106,0.0483)--(1.2206,0.0460)--(1.2305,0.0439)--(1.2405,0.0419)--(1.2505,0.0400)--(1.2605,0.0382)--(1.2705,0.0365)--(1.2804,0.0349)--(1.2904,0.0334)--(1.3004,0.0320)--(1.3104,0.0306)--(1.3204,0.0294)--(1.3304,0.0282)--(1.3403,0.0270)--(1.3503,0.0259)--(1.3603,0.0249)--(1.3703,0.0240)--(1.3803,0.0230)--(1.3902,0.0222)--(1.4002,0.0213)--(1.4102,0.0205)--(1.4202,0.0198)--(1.4302,0.0191)--(1.4401,0.0184)--(1.4501,0.0177)--(1.4601,0.0171)--(1.4701,0.0165)--(1.4801,0.0159)--(1.4901,0.0154)--(1.5000,0.0149)--(1.5100,0.0144)--(1.5200,0.0139);
\draw[blue!65!black,thick] (0.0228,0.0001)--(0.0328,0.0001)--(0.0428,0.0002)--(0.0527,0.0003)--(0.0627,0.0005)--(0.0727,0.0006)--(0.0827,0.0008)--(0.0927,0.0010)--(0.1027,0.0013)--(0.1126,0.0015)--(0.1226,0.0018)--(0.1326,0.0022)--(0.1426,0.0025)--(0.1526,0.0029)--(0.1625,0.0033)--(0.1725,0.0038)--(0.1825,0.0043)--(0.1925,0.0048)--(0.2025,0.0054)--(0.2124,0.0061)--(0.2224,0.0067)--(0.2324,0.0075)--(0.2424,0.0083)--(0.2524,0.0091)--(0.2624,0.0100)--(0.2723,0.0110)--(0.2823,0.0121)--(0.2923,0.0132)--(0.3023,0.0144)--(0.3123,0.0157)--(0.3222,0.0171)--(0.3322,0.0187)--(0.3422,0.0203)--(0.3522,0.0221)--(0.3622,0.0240)--(0.3721,0.0260)--(0.3821,0.0283)--(0.3921,0.0307)--(0.4021,0.0332)--(0.4121,0.0361)--(0.4221,0.0391)--(0.4320,0.0424)--(0.4420,0.0460)--(0.4520,0.0499)--(0.4620,0.0541)--(0.4720,0.0588)--(0.4819,0.0638)--(0.4919,0.0693)--(0.5019,0.0753)--(0.5119,0.0818)--(0.5219,0.0889)--(0.5318,0.0967)--(0.5418,0.1052)--(0.5518,0.1146)--(0.5618,0.1247)--(0.5718,0.1358)--(0.5818,0.1478)--(0.5917,0.1609)--(0.6017,0.1751)--(0.6117,0.1904)--(0.6217,0.2068)--(0.6317,0.2242)--(0.6416,0.2427)--(0.6516,0.2619)--(0.6616,0.2819)--(0.6716,0.3021)--(0.6816,0.3222)--(0.6915,0.3418)--(0.7015,0.3603)--(0.7115,0.3771)--(0.7215,0.3917)--(0.7315,0.4034)--(0.7415,0.4119)--(0.7514,0.4168)--(0.7614,0.4181)--(0.7714,0.4159)--(0.7814,0.4103)--(0.7914,0.4017)--(0.8013,0.3907)--(0.8113,0.3778)--(0.8213,0.3634)--(0.8313,0.3480)--(0.8413,0.3322)--(0.8513,0.3161)--(0.8612,0.3002)--(0.8712,0.2846)--(0.8812,0.2696)--(0.8912,0.2551)--(0.9012,0.2414)--(0.9111,0.2283)--(0.9211,0.2160)--(0.9311,0.2045)--(0.9411,0.1937)--(0.9511,0.1835)--(0.9610,0.1741)--(0.9710,0.1652)--(0.9810,0.1569)--(0.9910,0.1492)--(1.0010,0.1420)--(1.0110,0.1353)--(1.0209,0.1290)--(1.0309,0.1231)--(1.0409,0.1176)--(1.0509,0.1125)--(1.0609,0.1076)--(1.0708,0.1031)--(1.0808,0.0988)--(1.0908,0.0948)--(1.1008,0.0911)--(1.1108,0.0876)--(1.1207,0.0842)--(1.1307,0.0811)--(1.1407,0.0781)--(1.1507,0.0753)--(1.1607,0.0726)--(1.1707,0.0701)--(1.1806,0.0677)--(1.1906,0.0655)--(1.2006,0.0633)--(1.2106,0.0613)--(1.2206,0.0594)--(1.2305,0.0575)--(1.2405,0.0558)--(1.2505,0.0541)--(1.2605,0.0525)--(1.2705,0.0510)--(1.2804,0.0495)--(1.2904,0.0481)--(1.3004,0.0468)--(1.3104,0.0455)--(1.3204,0.0443)--(1.3304,0.0432)--(1.3403,0.0420)--(1.3503,0.0410)--(1.3603,0.0399)--(1.3703,0.0389)--(1.3803,0.0380)--(1.3902,0.0371)--(1.4002,0.0362)--(1.4102,0.0353)--(1.4202,0.0345)--(1.4302,0.0337)--(1.4401,0.0330)--(1.4501,0.0322)--(1.4601,0.0315)--(1.4701,0.0308)--(1.4801,0.0302)--(1.4901,0.0296)--(1.5000,0.0289)--(1.5100,0.0283)--(1.5200,0.0278);
\node[font=\scriptsize,anchor=south west] at (0,1.04) {$\operatorname{Re}\mathcal Y_{11}(i\omega)/Y_0$};
\node[font=\scriptsize] at (.8,-.18) {$\omega/\omega_0$};
\node[font=\scriptsize,anchor=east] at (0,0) {$0$};
\end{scope}
\end{scope}
\end{tikzpicture}
\caption{Failures of susceptibility-based admissibility tests. (a) Normalized minimum eigenvalue of $\operatorname{Herm}\tilde Y_{\rm ee}$ for Eq.~\eqref{eq:dielectric_counterexample_assignment}: $-|a_++a_--2|/8$. The white circle denotes vacuum, the dashed diagonal identical media, and the blue boundaries vacuum--substrate interfaces. (b) Full and dipolar admittances of Eq.~\eqref{eq:te_truncation_admittances} along the positive real Laplace axis, with $\gamma/\omega_0=0.4$, $c|k|/\omega_0=1$, and $Y_0=k^2F_0/(2\mu_0c^2\omega_0)$. The dipolar truncation violates temporal passivity despite nonnegative monochromatic power (inset). The dipolar curve is clipped at the lower boundary.}
\label{fig:admissibility_examples}
\end{figure*}

Their temporal passivity properties, however, are fundamentally different. The full admittance is positive real because $\operatorname{Re}[s a(s)]>0$ for $\operatorname{Re}s>0$ (Appendix~\ref{app:vector_family}). In contrast, the uncancelled dipolar term has a negative residue at $s=0$, and $\mathcal Y_{\rm dip}$ has a negative diagonal entry for every positive real $s$. It therefore violates Eq.~\eqref{eq:positive_real}. Figure~\ref{fig:admissibility_examples}(b) shows this failure despite nonnegative absorbed power on the regular frequency axis.

The quadrupole in this example is thus essential not merely to the accuracy of the response but to its temporal passivity. More generally, multipolar truncation requires a renewed assessment of physical admissibility; neither the admissibility of the full model nor monochromatic power tests suffice to guarantee that of the truncated model.

\subsection{Construction of physically admissible sheet models}\label{sec:constructive_completion}

The preceding examples show that intrinsic tensor restrictions can fail in dielectric embeddings and that multipolar truncation can destroy temporal passivity. We therefore reverse the modeling procedure: rather than assigning susceptibilities and testing their admissibility, we prescribe an admissible effective admittance and construct a compatible multipolar susceptibility model.

\paragraph{Admissibility by construction.}
For a prescribed admittance $Y$, substituting $j=Yx$ into Eq.~\eqref{eq:constitutive_feedback} gives
\begin{equation}
 L\boldsymbol{\chi}(A+BY)=Y .
 \label{eq:admittance_realization}
\end{equation}
Unlike the forward map in Eq.~\eqref{eq:effective_admittance}, this inverse realization condition is linear in the susceptibilities. A solution must also yield a regular constitutive closure.

In the Laplace domain, let $\mathcal Y_\star(s,\mathbf k)$ be a causal, reciprocal, positive-real admittance satisfying the reality condition of Sec.~\ref{sec:admissibility}. We seek a single dispersive susceptibility assignment realizing this target throughout the right half-plane, including for unequal dielectric half-spaces.

The construction uses two properties of the multipolar model. First, a constant selector $P_t$ extracts the average tangential fields from the excitation vector, independently of the embedding:
\begin{equation}
 P_t\mathcal A=I_4,\qquad P_t\mathcal B=0.
 \label{eq:synthesis_selector}
\end{equation}
Second, the ten independent symmetric traceless quadrupolar coordinates alone suffice to generate all four equivalent sheet currents. If $E_q$ embeds these coordinates in the full moment space, the GSTCs admit a right inverse $R_q$ satisfying
\begin{equation}
 \mathcal L(s,\mathbf k)E_qR_q(s,\mathbf k)=I_4,
 \qquad s\ne0.
 \label{eq:synthesis_current_inverse}
\end{equation}
An explicit right inverse, including at normal incidence, is derived in Appendix~\ref{app:quadrupolar_inverse}. Together these properties give
\begin{equation}
 \boxed{\boldsymbol{\chi}_\star(s,\mathbf k)
 =E_qR_q(s,\mathbf k)\mathcal Y_\star(s,\mathbf k)P_t.}
 \label{eq:feedback_free_synthesis}
\end{equation}
Since $\boldsymbol{\chi}_\star\mathcal B=0$ and $\mathcal L\boldsymbol{\chi}_\star\mathcal A=\mathcal Y_\star$, the feedback and direct current matrices are
\begin{equation}
 \mathcal U_\star=I_4,\qquad
 \mathcal V_\star=\mathcal Y_\star,
 \label{eq:synthesis_feedback}
\end{equation}
and the realized admittance is exactly
\begin{equation}
 \mathcal Y[\boldsymbol{\chi}_\star]=\mathcal Y_\star.
 \label{eq:synthesized_response}
\end{equation}
The construction therefore guarantees both closure regularity and the prescribed physical properties of the response, independently of the isotropic dielectric embedding.

Appendix~\ref{app:quadrupolar_inverse} establishes analyticity, reality symmetry and preservation of causality at fixed real momentum. The right inverse contains $1/s$ for nonzero momentum and $1/s^2$ at normal incidence; static regularity and uniformity near zero momentum require separate consideration.

\paragraph{Constitutive freedom.}
The physical response can be fixed without uniquely fixing its constituent moments. In particular,
\begin{equation}
 \boldsymbol{\chi}=\boldsymbol{\chi}_\star+N,
 \qquad \mathcal L N=0
 \label{eq:synthesis_nullspace_family}
\end{equation}
preserves both the admittance and closure regularity for compatible additions $N$ satisfying the required causal and reality conditions. The ten-dimensional quadrupolar moment space has a six-dimensional current nullspace for $s\ne0$.

Tangential dipoles alone can also span the four currents. The significance of higher moments here is that they provide additional constitutive freedom, rather than being necessary for realizing every admissible target. Appendix~\ref{app:prescribed_dipolar} gives the residual-current completion for prescribed dipolar rows, subject to compatibility and regularity. Thus distinct moment assignments can realize the same response.

\begin{figure*}[!t]
 \centering
 \resizebox{\textwidth}{!}{\input{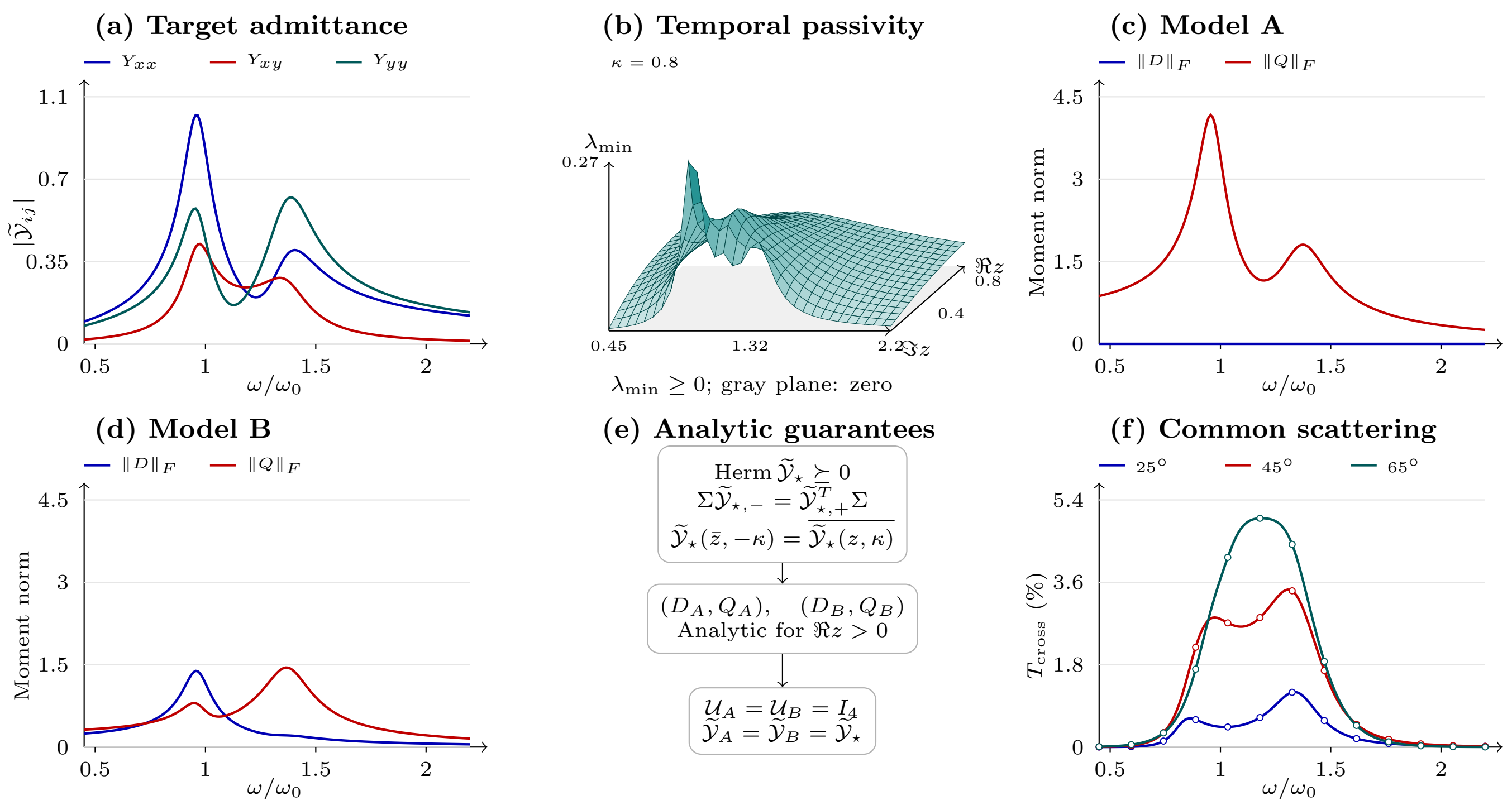}}
 \caption{Admissibility by construction for a two-resonance anisotropic sheet. (a) Magnitudes of selected electric admittance entries. (b) Minimum eigenvalue of the Hermitian part of the admittance in the right half-plane region shown (gray base: zero). (c,d) Frobenius norms of the dimensionless dipolar and quadrupolar maps for Models A and B, respectively, on identical scales. Panels (a)--(d) use $\kappa=0.8$. (e) Admissibility and realization identities; subscripts $+$ and $-$ denote opposite momenta. (f) TM-to-TE transmitted power fraction versus frequency for three incidence angles from vacuum into a dielectric of index $1.5$. Solid curves (Model A) and open markers (Model B) coincide. All curves are evaluations of the analytical model.}
 \label{fig:admissible_synthesis}
\end{figure*}

\paragraph{Broadband synthesis example.}
Consider an anisotropic sheet with two spatially dispersive Lorentz resonances and different polarization axes. In the normalized field and current coordinates of Appendix~\ref{app:normalization}, prescribe
\begin{align}
 \widetilde{\mathcal Y}_\star(z,\kappa)
 &=Y_1(z,\kappa)+Y_2(z,\kappa),\\
 Y_r(z,\kappa)
 &=\frac{a(\kappa)z}{z^2+\gamma_r z+\Omega_r^2(\kappa)}
 \begin{pmatrix}F_{{\rm e}r}&0\\0&F_{{\rm m}r}\end{pmatrix},
 \label{eq:synthesis_lorentz_kernel}
\end{align}
where $z=s/\omega_0$ is the normalized Laplace variable and $\kappa=k_x/(\omega_0/c)$ for $\mathbf k=k_x\hat{\mathbf x}$. With positive-definite real symmetric residues, positive damping, positive resonance frequencies and even momentum dependence with $a(\kappa)\geq0$, the target is reciprocal, causal and positive real. Appendix~\ref{app:illustrative_synthesis} gives the parameters and verifies admissibility. The noncommuting electric residues have different principal axes, giving distinct polarization coupling across the two resonance bands [Fig.~\ref{fig:admissible_synthesis}(a,b)].

We construct two realizations: Model A uses quadrupoles alone, while Model B assigns $85\%$ of the first resonance and $15\%$ of the second to tangential dipoles, with quadrupoles supplying the remaining current. Both use only average tangential field excitations. The explicit assignments [Eq.~\eqref{eq:illustrative_realizations}] give susceptibilities that are regular at $s=0$ and extend continuously to zero at zero momentum in this example. They have different moment spectra [Fig.~\ref{fig:admissible_synthesis}(c,d)] but exactly the same admittance [Eq.~\eqref{eq:illustrative_equal_response}].

The common admittance produces identical scattering wherever the exterior problem is well posed. Figure~\ref{fig:admissible_synthesis}(f) illustrates this for TM-to-TE transmission into a dielectric substrate, using the scattering calculation in Appendix~\ref{app:illustrative_synthesis}. Although scattering depends on the embedding, the two constitutive realizations remain equivalent and physically admissible within the adopted sheet model.

The synthesis thus separates the selection of a physically admissible response from the choice of its multipolar realization. Physical admissibility is guaranteed by construction, while the distribution of the response among the moments remains a degree of freedom.

\section{Conclusion}\label{sec:conclusion}

We have established a response-based framework for the physical admissibility of dispersive multipolar electromagnetic sheet models. The effective surface admittance retains the general dipolar--quadrupolar constitutive closure, including the feedback associated with unequal dielectric half-spaces. Expressing the sheet-level power and Lorentz reaction identities in terms of this admittance identifies the physical constraints independently of the particular moment representation, while the Laplace-domain formulation distinguishes temporal passivity from monochromatic power balance. These constraints yield exact conditions on the susceptibilities for regular constitutive closures.

Two consequences of this distinction are of particular physical significance. Conventional susceptibility restrictions that suffice in vacuum need not ensure reciprocity and losslessness in a dielectric embedding, even for susceptibilities assigned specifically to that environment. Furthermore, higher multipolar moments can be essential to the temporal passivity of a given constitutive representation, rather than merely improving its scattering accuracy. The admissibility of a reduced representation must therefore be established by preserving the physical response or reassessing its admissibility.

The constructive inverse formulation provides an alternative to imposing and testing susceptibility restrictions. A prescribed causal, reciprocal, positive-real admittance satisfying the reality condition can be realized by dispersive multipolar susceptibilities at fixed real tangential momentum throughout the open right half-plane. The construction guarantees closure regularity through identity constitutive feedback, independently of the isotropic dielectric embedding. The quadrupolar current nullspace and the freedom to allocate currents among dipolar and quadrupolar moments yield nonunique constitutive realizations. In the broadband example, substantially different susceptibility spectra produce identical admittances and polarization-conversion spectra.

The joint spatial and temporal dispersion of the effective admittance makes established methods of passive linear-response theory and network synthesis~\cite{Anderson1973} applicable to the general multipolar closure considered here. Together with power-normalized scattering descriptions~\cite{Kurokawa1965}, this provides a basis for investigating broadband response bounds, synthesis, and scattering poles and reflection zeros~\cite{Sweeney2020}, while exploiting the freedom in the constitutive realization. Extension to multipolar orders beyond the quadrupoles is a natural next step: the response-level power and reaction conditions remain applicable when expressed in equivalent currents, whereas the higher-order source and excitation maps and their constitutive realization require further derivation. The response can thus be selected to satisfy the desired physical requirements before its constitutive realization is determined, yielding dispersive multipolar sheet models that are physically admissible by construction.

\section*{AI Use Disclosure}
ChatGPT (OpenAI, GPT-6.2) and AI-assisted coding tools were used to assist with analytical derivations and verification, manuscript editing, analytical figure code, and formalization of mathematical results in Lean~4. Verification includes machine checking of the encoded mathematical statements. The authors remain responsible for the derivations, their correspondence to the physical model, and the final scientific content.

\appendix
\section{Supporting derivations for the admissibility framework}\label{app:excitation}

\subsection{Constitutive normalization and excitation maps}\label{app:normalization}

To express the normalization of Eq.~\eqref{eq:published_closure}~\cite{Tiukuvaara2023,Achouri2023Multipolar} in matrix form, set $\tilde x=N_x x$, $\tilde j=N_j j$, where $N_x=\operatorname{diag}(I_2,\eta_0 I_2)$ and $N_j=\operatorname{diag}(\eta_0 I_2,I_2)$. In the excitation order $(E,H,\nabla E,\nabla H)$, with matching tensor vectorization, define
\begin{align}
 N_m&=\operatorname{diag}(\epsilon_0^{-1}I_3,\eta_0 I_3,
 k_0\epsilon_0^{-1}I_9,k_0\eta_0 I_9),\\
 N_f&=\operatorname{diag}(I_3,\eta_0 I_3,
 (2k_0)^{-1}I_9,\eta_0(2k_0)^{-1}I_9).
\end{align}
Then $\tilde m=N_m m$, $\tilde f=N_f f$, and
\begin{equation}
 X=N_m\;\boldsymbol{\chi}\;N_f^{-1}.
 \label{eq:normalized_susceptibility}
\end{equation}
Following Eq.~(8) of Ref.~\cite{Tiukuvaara2023}, weight the normal acting values by $E_n\mapsto a E_n$, $H_n\mapsto b H_n$, with $a=\epsilon/\epsilon_0$ and $b=\mu/\mu_0$; leave the physical gradient entries unweighted.

For each homogeneous isotropic half-space, the source-free Maxwell equations~\cite{Harrington2001} give the following two-polarization extension of Eq.~(10) of Ref.~\cite{Tiukuvaara2023}. With $(\mathbf k\times\mathbf H_\parallel)_n=\hat{\mathbf n}\cdot(\mathbf k\times\mathbf H_\parallel)$ and $\omega\ne0$,
\begin{align}
 E_n&=\frac{(\mathbf k\times\mathbf H_\parallel)_n}{\omega\epsilon},
 &
 H_n&=-\frac{(\mathbf k\times\mathbf E_\parallel)_n}{\omega\mu},\\
 \partial_z\mathbf E_\parallel
 &=i\mathbf k E_n+i\omega\mu R\mathbf H_\parallel,\\
 \partial_z\mathbf H_\parallel
 &=i\mathbf k H_n-i\omega\epsilon R\mathbf E_\parallel,\\
 \partial_z E_n&=-i\mathbf k^T\mathbf E_\parallel,
 &
 \partial_z H_n&=-i\mathbf k^T\mathbf H_\parallel .
\end{align}
Together with $\nabla_\parallel=i\mathbf k$, the above relations followed by acting-value weighting and normalization define $G_{a,b}\in\mathbb C^{24\times4}$ for arbitrary tangential traces.

\subsection{Embedding-dependent feedback and admittance}\label{app:feedback}

Averaging the one-sided excitation vectors~\cite{Kuester2003,Tiukuvaara2023} yields $\tilde f=C\tilde x+D_{\rm j}\Delta\tilde x$, where
\begin{align}
 C&=\tfrac12(G_{a_+,b_+}+G_{a_-,b_-}),\\
 D_{\rm j}&=\tfrac14(G_{a_+,b_+}-G_{a_-,b_-}).
\end{align}
The jump of the normalized tangential field vector follows from Eq.~\eqref{eq:equivalent_currents}~\cite{BudhuGrbic2021,Szymanski2021}: $\Delta\tilde x=H_{\rm j}\tilde j$, where
\begin{equation}
 H_{\rm j}=\begin{pmatrix}0&R\\-R&0\end{pmatrix}.
\end{equation}
Thus $\tilde f=C\tilde x+B_{\rm n}\tilde j$, with $B_{\rm n}=D_{\rm j}H_{\rm j}$. Hence, in physical coordinates,
\begin{equation}
 A=N_f^{-1}CN_x,\qquad B=N_f^{-1}B_{\rm n}N_j.
 \label{eq:physical_excitation_maps}
\end{equation}
With $F=N_j L N_m^{-1}$ and $\tilde Y=N_j YN_x^{-1}$, the normalized current equation is $(I_4-FXB_{\rm n})\tilde j=FXC\tilde x$.

For a regular closure, the normalized counterpart of Eq.~\eqref{eq:effective_admittance} is
\begin{equation}
 \tilde Y=(I_4-FXD_{\rm j}H_{\rm j})^{-1}FXC .
 \label{eq:dielectric_admittance}
\end{equation}
In identical media $D_{\rm j}=0$, $C=G_{a,b}$, and $\tilde Y=FXG_{a,b}$.

\subsection{Power and reaction pairings}\label{app:flux_pairings}

For the standard power and Lorentz reaction balances~\cite{Harrington2001,Achouri2021}, the required midpoint--jump identity for two field states $a,b$ is
\begin{align}
 &\hat{\mathbf n}\cdot
 (\mathbf E_a^-\times\mathbf H_b^-
 -\mathbf E_a^+\times\mathbf H_b^+) \nonumber\\
 &\quad=\overline{\mathbf E}_{a,\parallel}\cdot\mathbf J_b
 +\mathbf K_a\cdot\overline{\mathbf H}_{b,\parallel}.
\end{align}
Taking $b$ as the complex-conjugated field state of $a$ and taking half the real part gives Eq.~\eqref{eq:surface_power}; antisymmetrizing in $a,b$ gives Eq.~\eqref{eq:surface_reaction}.

Since $N_j=\eta_0 N_x^{-1}$ and $N_x\Sigma=\Sigma N_x$,
\begin{align}
 \operatorname{Herm}\tilde Y
 &=\eta_0 N_x^{-1}\operatorname{Herm}(Y)N_x^{-1},\\
 \Sigma\tilde Y_--\tilde Y_+^T\Sigma
 &=\eta_0 N_x^{-1}(\Sigma Y_--Y_+^T\Sigma)N_x^{-1}.
\end{align}
The subscripts denote opposite momenta. Invertible congruence~\cite{HornJohnson2013} thus preserves both admissibility tests.

\subsection{Constitutive admissibility tests}\label{app:constitutive_tests}

For invertible $U$ in Eq.~\eqref{eq:closure_UV}, $Y=U^{-1}V$ gives
\begin{equation}
 U(Y+Y^\dagger)U^\dagger=VU^\dagger+UV^\dagger.
\end{equation}
By the standard congruence property~\cite{HornJohnson2013}, this yields Eq.~\eqref{eq:chi_power}, including equality for losslessness.

Likewise the reciprocity residual satisfies
\begin{align}
 &U_-\Sigma(\Sigma Y_--Y_+^T\Sigma)\Sigma U_+^T\nonumber\\
 &\quad=V_-\Sigma U_+^T-U_-\Sigma V_+^T.
\end{align}
This follows from $\Sigma^2=I_4$ and $Y_\pm=U_\pm^{-1}V_\pm$. The invertible factors preserve the vanishing of the residual, proving Eq.~\eqref{eq:chi_reciprocity}. The same identities apply in Laplace coordinates.

\section{Embedding dependence and multipolar truncation}\label{app:applications}

\subsection{Vacuum adjoint identity and dielectric conditions}

Use the normalized coordinates and excitation map of Appendix~\ref{app:normalization}, with $G=G_{1,1}$, $t=R\mathbf k$, and $\tilde m=(p,m,q,h_q)^T$ denoting the four normalized moment blocks. Direct contraction with the vacuum excitation map gives
\begin{align}
 ik_0(G^\dagger\tilde m)_E
 &=ik_0p_\parallel-i m_n t
 -\tfrac12\mathbf a_q
 -\tfrac{b_{h_q}}{2k_0}t+\tfrac{k_0}{2}R(h_q)_{\parallel n},\\
 ik_0(G^\dagger\tilde m)_H
 &=ik_0m_\parallel+i p_n t
 -\tfrac12\mathbf a_{h_q}
 +\tfrac{b_q}{2k_0}t-\tfrac{k_0}{2}Rq_{\parallel n}.
\end{align}
Comparison with the normalized GSTC currents of Eqs.~\eqref{eq:gstc_electric} and \eqref{eq:gstc_magnetic} proves Eq.~\eqref{eq:vacuum_source_adjoint} without imposing tensor symmetry.

The momentum and conjugation parity of the GSTC source blocks~\cite{Tiukuvaara2023,Achouri2023Multipolar} gives
\begin{equation}
 L(-\mathbf k)^T\Sigma=D L(\mathbf k)^\dagger,
 \label{eq:momentum_conjugacy}
\end{equation}
with $D$ as in Eq.~\eqref{eq:vacuum_sufficient} and its accompanying definition. The real diagonal normalization of Appendix~\ref{app:normalization} preserves this identity for $F$.

Applying the normalized power and reaction tests (Appendix~\ref{app:flux_pairings}) to Eq.~\eqref{eq:vacuum_admittance} gives, for $\omega>0$,
\begin{equation}
 G^\dagger\frac{X^\dagger-X}{2i}G\succeq0 ,
 \label{eq:vacuum_passivity}
\end{equation}
with equality for losslessness. Equation~\eqref{eq:momentum_conjugacy} gives the exact reciprocity test
\begin{equation}
 G^T[DX(-\mathbf k)D-X(\mathbf k)^T]\overline G=0 .
 \label{eq:vacuum_reciprocity}
\end{equation}
Here $\overline G$ denotes entrywise conjugation. For identical dielectric media the same tests, with $\tilde Y=ik_0G^\dagger XG_{a,b}$, give
\begin{align}
 G^\dagger XG_{a,b}&=G_{a,b}^\dagger X^\dagger G,\\
 \Sigma G_-^\dagger X_-G_{a,b,-}
 &=(G^\dagger XG_{a,b})^T\Sigma,
 \label{eq:dielectric_exact_conditions}
\end{align}
where the minus subscript denotes momentum reversal.

\subsection{Counterexample at an unequal dielectric interface}\label{app:unequal_dielectric_example}

For Eq.~\eqref{eq:dielectric_counterexample_assignment} at normal incidence with $b_\pm=1$, set $\tilde m=\lambda v\xi$, $\xi=v^\dagger\tilde f$. The normalized GSTCs give
\begin{equation}
 Fv=ik_0(1,1/2,0,0)^T .
\end{equation}
Hence $\mathbf K=0$ and $\Delta\mathbf E_\parallel=0$. The one-sided Maxwell relation of Appendix~\ref{app:normalization} yields
\begin{equation}
 \partial_z H_x^\pm=i\omega\epsilon_\pm E_y^\pm .
\end{equation}
With $\partial_x H_z=0$,
\begin{align}
 \xi&=\overline E_x
 -i\frac{\eta_0}{2k_0}\overline{\partial_z H_x}\\
 &=\overline E_x+\frac{\bar a}{2}\overline E_y.
\end{align}
Substitution into $\tilde j=\lambda Fv\xi$ gives the admittance of Sec.~\ref{sec:dielectric_failure}, with all other blocks zero. Moreover, $FXB_{\rm n}$ maps magnetic current coordinates into electric current coordinates only. Thus $(FXB_{\rm n})^2=0$ and $FXB_{\rm n}FXC=0$, proving closure regularity and the exact cancellation of feedback in Eq.~\eqref{eq:dielectric_admittance}.

\subsection{TE reduction in vacuum}\label{app:te_reduction}

For the TE sector with $\mathbf k=k\hat{\mathbf x}$, the published multipolar GSTCs~\cite{Tiukuvaara2023,Achouri2023Multipolar} specialize to
\begin{align}
 J_y={}&sP_y-ikM_z-\tfrac{isk}{2}Q_{xy}\nonumber\\
 &-\left(k^2+\tfrac{s^2}{2c^2}\right)S_{xz},\\
 K_x={}&s\mu_0 M_x+\tfrac{isk\mu_0}{2}(S_{zz}-S_{xx})
 -\tfrac{s^2\mu_0}{2}Q_{yz},
 \label{eq:te_multipolar_gstcs}
\end{align}
for symmetric quadrupoles. Setting $Q=S=0$ and $P_y^{\rm eff}=P_y-ikM_z/s$ gives the tangential dipolar representation~\cite{Phaneuf2026},
\begin{equation}
 \binom{P_y^{\rm eff}}{\mu_0 M_x}
 =C_{\rm TE}(s,k)\binom{\overline E_y}{\overline H_x},
 \qquad \mathcal Y_{\rm TE}=s C_{\rm TE}.
\end{equation}
In the multipolar representation, the increments $\delta M_z=s\beta\overline E_y$ and $\delta Q_{xy}=\delta Q_{yx}=-2\beta\overline E_y$ cancel in Eq.~\eqref{eq:te_multipolar_gstcs}, for any real causal scalar kernel $\beta$ with sufficient decay. Thus the corresponding normal magnetoelectric coefficient need not vanish in every admissible multipolar representation.

\subsection{Temporal passivity of the full and truncated responses}\label{app:vector_family}

For the Lorentz kernel of Sec.~\ref{sec:multipolar_extension},
\begin{equation}
 \operatorname{Re}[s a(s)]=
 F_0\frac{\sigma(|s|^2+\omega_0^2)+\gamma|s|^2}
 {|s^2+\gamma s+\omega_0^2|^2}>0,\quad
 \sigma=\operatorname{Re}s>0 .
\end{equation}
This proves positive-realness of the full TE admittance in Eq.~\eqref{eq:te_truncation_admittances}.

For a vector generalization in vacuum, let $T=\operatorname{diag}(R\mathbf k,R\mathbf k)\in\mathbb R^{4\times2}$ and use the normal-field relations of Appendix~\ref{app:normalization}:
\begin{equation}
 g=\binom{\mu_0\overline H_n}{-\epsilon_0\overline E_n}
 =-\frac{i}{s}T^Tx,\quad
 D_c=\operatorname{diag}\left(\frac{\epsilon_0\mu_0}{2},-\frac{\mu_0}{2}\right).
\end{equation}
For a $2\times2$ dispersive kernel $\mathcal C(s,\mathbf k)$, take $u=D_c^{-1}\mathcal Cg$ and
\begin{align}
 \binom{M_n}{P_n}&=|\mathbf k|^2u,\\
 (S_{\parallel n},Q_{\parallel n})&=-i\mathbf k u^T .
 \label{eq:vector_completion_moments}
\end{align}
Complete the normal rows by symmetry; all other moments vanish. Then $\mathbf a_Q=\mathbf a_S=0$, $b_S=-2i|\mathbf k|^2u_1$, and $b_Q=-2i|\mathbf k|^2u_2$. The normal dipole terms cancel the contraction terms in the GSTCs, leaving $j=is^2TD_cu$. Deleting the quadrupoles instead gives $j=-i|\mathbf k|^2T\operatorname{diag}(1,-1/\epsilon_0)u$. Since $\operatorname{diag}(1,-1/\epsilon_0)D_c^{-1}=2c^2I_2$,
\begin{equation}
 \mathcal Y_{\rm multi}=sT\mathcal CT^T,\qquad
 \mathcal Y_{\rm dip}=-\frac{2c^2|\mathbf k|^2}{s}T\mathcal CT^T .
 \label{eq:vector_completion_admittance}
\end{equation}

Suppose $\mathcal C$ is analytic in the right half-plane, regular at $s=0$, and satisfies the causal-kernel growth bounds and
\begin{align}
 \mathcal C(\bar s,-\mathbf k)&=\overline{\mathcal C(s,\mathbf k)},\\
 \Sigma_2\mathcal C(s,-\mathbf k)&=\mathcal C(s,\mathbf k)^T\Sigma_2,\\
 \operatorname{Herm}[s\mathcal C(s,\mathbf k)]&\succeq0,
 \quad \operatorname{Re}s>0 ,
 \label{eq:vector_completion_conditions}
\end{align}
where $\Sigma_2=\operatorname{diag}(1,-1)$. By $T(-\mathbf k)=-T(\mathbf k)$, $\Sigma T=T\Sigma_2$, and congruence~\cite{HornJohnson2013}, the full response satisfies the reality, reciprocity and positive-real tests of Sec.~\ref{sec:admissibility}. The physical susceptibility in Eq.~\eqref{eq:vector_completion_moments} has no inverse frequency or momentum factors. The TE example uses $\mathcal C=\operatorname{diag}(a/(2\mu_0c^2),0)$.

At regular nonzero real frequencies, the two admittances differ by the positive factor $2c^2|\mathbf k|^2/\omega^2$. At positive real $s$ the factor is negative for $\mathbf k\ne0$; a nonzero Hermitian part of the full response there implies failure of temporal passivity for the truncation.

\section{Derivations for admissible sheet synthesis}\label{app:synthesis}

The target $\mathcal Y_\star$ satisfies the conditions of Sec.~\ref{sec:admissibility}.

\subsection{Explicit quadrupolar current right inverse}\label{app:quadrupolar_inverse}

At fixed real physical momentum $\mathbf k=(k_x,k_y)^T\ne0$, define
\begin{equation}
 D_k=
 \begin{pmatrix}k_x&k_y\\-k_y&k_x\end{pmatrix},
 \qquad D_k^TD_k=|\mathbf k|^2I_2 .
\end{equation}
Choose the symmetric traceless tensors with only the in-plane blocks
\begin{equation}
 Q_\parallel=\begin{pmatrix}q_1&q_2\\q_2&-q_1\end{pmatrix},
 \quad S_\parallel=\begin{pmatrix}h_1&h_2\\h_2&-h_1\end{pmatrix}.
\end{equation}
All other moments vanish. With $s=i\omega$ on the harmonic axis, the Laplace continuation of Eqs.~\eqref{eq:gstc_electric} and \eqref{eq:gstc_magnetic}~\cite{Tiukuvaara2023,Achouri2023Multipolar} gives
\begin{equation}
 \mathbf J=-\frac{is}{2}D_k\binom{q_1}{q_2},\qquad
 \mathbf K=-\frac{is\mu_0}{2}D_k\binom{h_1}{h_2}.
\end{equation}
For any prescribed current vector, take
\begin{equation}
 \binom{q_1}{q_2}=\frac{2iD_k^T\mathbf J}{s|\mathbf k|^2},
 \qquad
 \binom{h_1}{h_2}=\frac{2iD_k^T\mathbf K}{s\mu_0|\mathbf k|^2}.
\end{equation}
At normal incidence instead choose only the mixed normal--tangential components (with the transposed components equal). The GSTCs give
\begin{equation}
 \mathbf J=-\frac{s^2\epsilon_0\mu_0}{2}R S_{\parallel n},
 \quad
 \mathbf K=\frac{s^2\mu_0}{2}R Q_{\parallel n}.
\end{equation}
Since $R^2=-I_2$, the corresponding right inverse is
\begin{equation}
 S_{\parallel n}=\frac{2R\mathbf J}{s^2\epsilon_0\mu_0},
 \qquad
 Q_{\parallel n}=-\frac{2R\mathbf K}{s^2\mu_0}.
\end{equation}
These constructions prove Eq.~\eqref{eq:synthesis_current_inverse}. At fixed physical momentum, the right inverse in these coordinates has the form
\begin{equation}
 R_q(s,\mathbf k)=
 \begin{cases}
 C_q(\mathbf k)/s,&\mathbf k\ne0,\\
 C_q(0)/s^2,&\mathbf k=0,
 \end{cases}
 \label{eq:synthesis_frequency_factors}
\end{equation}
where $C_q$ is frequency independent, imaginary and odd in momentum for $\mathbf k\ne0$, and real at normal incidence. Thus $R_q(\bar s,-\mathbf k)=\overline{R_q(s,\mathbf k)}$, and Eq.~\eqref{eq:feedback_free_synthesis} inherits the target's reality symmetry. Both branches are analytic for $\operatorname{Re}s>0$; their inverse powers of $s$ correspond to causal integration kernels $\Theta(t)$ and $t\Theta(t)$, where $\Theta$ is the unit step. The real, constant maps $E_q,P_t$ preserve these properties. If the target is polynomially bounded on $\operatorname{Re}s\geq\sigma>0$ at fixed $\mathbf k$, the susceptibility has the same growth bound up to the constant $\|E_q\|\|C_q(\mathbf k)\|\|P_t\|/\sigma^d$, where $d=1,2$ for the respective branches. No uniform bound as $\sigma\to0$ or $\mathbf k\to0$ is implied. Multiplying the physical quadrupolar output coordinates by $k_0\propto s$ removes one inverse power of $s$ in the normalized coordinates.

At $s=0$, the surviving quadrupolar currents are $-\tfrac12 b_S R\mathbf k$ and $\tfrac{b_Q}{2\epsilon_0}R\mathbf k$. For $\mathbf k\ne0$, the mixed components independently span both transverse currents through $b_T=2\mathbf k^TT_{\parallel n}$, giving a two-dimensional static range; both currents vanish at $\mathbf k=0$.

\subsection{Constitutive freedom with prescribed dipolar rows}\label{app:prescribed_dipolar}

With prescribed dipolar output rows $\boldsymbol{\chi}_d$ and adjustable quadrupolar rows $\boldsymbol{\chi}_q$ (both embedded in the full moment space), the inverse realization condition~\eqref{eq:admittance_realization} of Sec.~\ref{sec:constructive_completion} gives
\begin{align}
 &\mathcal L\boldsymbol{\chi}_q(\mathcal A+\mathcal B\mathcal Y_\star)
 \nonumber\\
 &\qquad=\mathcal Y_\star-
 \mathcal L\boldsymbol{\chi}_d(\mathcal A+\mathcal B\mathcal Y_\star).
 \label{eq:multipolar_completion}
\end{align}
A solution realizes the target wherever the closure is regular. In vacuum, $\mathcal B=0$ and $P_t\mathcal A=I_4$; using the right inverse in Appendix~\ref{app:quadrupolar_inverse}, an explicit solution for arbitrary prescribed dipolar rows is
\begin{equation}
 \boldsymbol{\chi}_q=
 E_qR_q(\mathcal Y_\star-\mathcal L\boldsymbol{\chi}_d\mathcal A)P_t,
 \qquad s\ne0 .
 \label{eq:quadrupolar_completion_explicit}
\end{equation}
The solution is nonunique by the current-nullspace argument in Sec.~\ref{sec:constructive_completion}. Broadband realization requires the stated causal and reality conditions. For unequal isotropic embeddings, the feedback-free construction of Eqs.~\eqref{eq:synthesis_selector}--\eqref{eq:synthesized_response} applies when the right inverse and excitation maps use the same physical coordinates; arbitrary fixed dipolar rows must instead satisfy the compatibility condition above and yield regular closure.

\subsection{Two-resonance anisotropic synthesis and scattering}
\label{app:illustrative_synthesis}

Use $k_{\rm r}=\omega_0/c$, $a(\kappa)=0.32\kappa^2/(1+\kappa^2)$, $\Omega_1^2=0.8^2+0.45\kappa^2$, $\Omega_2^2=1.35^2+0.10\kappa^2$, and $(\gamma_1,\gamma_2)=(0.13,0.22)$ in Eq.~\eqref{eq:synthesis_lorentz_kernel}.

The dimensionless target of Eq.~\eqref{eq:synthesis_lorentz_kernel} uses
\begin{align}
 F_{{\rm e}1}&=\begin{pmatrix}1&0.45\\0.45&0.55\end{pmatrix},
 &F_{{\rm e}2}&=\begin{pmatrix}0.55&-0.48\\-0.48&1.05\end{pmatrix},\\
 F_{{\rm m}1}&=\begin{pmatrix}0.8&-0.27\\-0.27&0.45\end{pmatrix},
 &F_{{\rm m}2}&=\begin{pmatrix}0.38&0.31\\0.31&0.72\end{pmatrix}.
\end{align}
All four residues are positive definite by the principal-minor criterion~\cite{HornJohnson2013}. The Lorentz positive-real identity of Appendix~\ref{app:vector_family} applies with $s$ replaced by $z$, $F_0=1$, $\gamma=\gamma_r$ and $\omega_0=\Omega_r(\kappa)$. Since $a(\kappa)\geq0$ and the poles lie strictly in the left half-plane, the target is positive real and causal~\cite{Anderson1973}. The symmetric block-diagonal residues commute with $\Sigma$; even momentum dependence and real coefficients give reciprocity and reality. The constant field and current normalization preserves these properties (Appendix~\ref{app:flux_pairings}).

Define the fixed-reference moment coordinates
\begin{align}
 p&=k_{\rm r}\mathbf P_\parallel/\epsilon_0,
 &m&=k_{\rm r}\eta_0\mathbf M_\parallel,\\
 (q_1,q_2)&=k_{\rm r}^2(Q_{xx},Q_{xy})/\epsilon_0,
 &(h_1,h_2)&=k_{\rm r}^2\eta_0(S_{xx},S_{xy}).
\end{align}
Take $Q_{yy}=-Q_{xx}$, $S_{yy}=-S_{xx}$ and symmetric off-diagonal entries, with all normal components zero. In the fixed Cartesian frame let $d=(p_x,p_y,m_x,m_y)^T$ and $q=(q_1,q_2,h_1,h_2)^T$. From the physical GSTCs with $\omega=-is$ and $k_x=k_{\rm r}\kappa$, the normalized source relation is $\tilde j=zd-iz\kappa q/2$. Define $d=D_{A,B}\tilde x$ and $q=Q_{A,B}\tilde x$ through
\begin{align}
 D_A&=0,& Q_A&=\frac{2i}{z\kappa}\widetilde{\mathcal Y}_\star,\\
 D_B&=\frac{Y_d}{z},&
 Q_B&=\frac{2i}{z\kappa}
       (\widetilde{\mathcal Y}_\star-Y_d),
 \label{eq:illustrative_realizations}\\
 Y_d&=0.85Y_1+0.15Y_2.\nonumber
\end{align}
Substitution gives
\begin{equation}
 zD_A-\frac{iz\kappa}{2}Q_A
 =zD_B-\frac{iz\kappa}{2}Q_B
 =\widetilde{\mathcal Y}_\star .
 \label{eq:illustrative_equal_response}
\end{equation}
Both assignments use only tangential fields, so Eq.~\eqref{eq:synthesis_selector} gives $\mathcal U_A=\mathcal U_B=I_4$. The target numerator cancels the apparent singularities at $z=0$ and $\kappa=0$. The dipolar maps are even in momentum with real rational coefficients, and the quadrupolar maps have imaginary odd momentum factors; hence both have causal, regular kernels and the required reality symmetry. Their continuous zero extension at $\kappa=0$ realizes the zero target.

For the scattering calculation, the lower and upper half-spaces are nonmagnetic with refractive indices $n_-=1$ and $n_+=1.5$. For a real positive frequency $w=\omega/\omega_0$ and incidence angle $\theta$ from the lower half-space, take $\kappa=w\sin\theta$. Write $C_\pm=\sqrt{1-\kappa^2/(n_\pm w)^2}$ and the normalized tangential wave-admittance matrices
\begin{equation}
 W_\pm=\operatorname{diag}(n_\pm/C_\pm,n_\pm C_\pm).
\end{equation}
For tangential incident, reflected and transmitted electric-field vectors $e,r,t$, respectively, the traces are
\begin{align}
 E^-&=e+r,& \eta_0 H^-&=R W_-(e-r),\\
 E^+&=t,& \eta_0 H^+&=R W_+t.
\end{align}
Substitution into $\tilde j=(R\eta_0\Delta H,-R\Delta E)^T$ and $\tilde j=\widetilde{\mathcal Y}\tilde x$ produces four linear equations for $(r,t)$, with $\tilde x$ the arithmetic mean of the two field traces. For TM incidence $e=(1,0)^T$, the power-normalized TM-to-TE transmittance~\cite{Kurokawa1965} is $(W_+)_{22}|t_y|^2/(W_-)_{11}$. In Fig.~\ref{fig:admissible_synthesis}(f), the admittance and scattering are evaluated separately for each moment assignment. The maximum absolute differences in admittance entries and scattering amplitudes across the sampled grid are below $10^{-14}$ in double precision, illustrating Eq.~\eqref{eq:illustrative_equal_response}.

\bibliography{references}

@article{Achouri2021,
  author = {Achouri, Karim and Martin, Olivier J. F.},
  title = {Extension of Lorentz reciprocity and Poynting theorems for spatially dispersive media with quadrupolar responses},
  journal = {Phys. Rev. B},
  volume = {104},
  pages = {165426},
  year = {2021},
  doi = {10.1103/PhysRevB.104.165426}
}

@article{Tiukuvaara2023,
  author = {Tiukuvaara, Ville and Martin, Olivier J. F. and Achouri, Karim},
  title = {Quadrupolar susceptibility modeling of substrated metasurfaces},
  journal = {Opt. Express},
  volume = {31},
  pages = {22982},
  year = {2023},
  doi = {10.1364/OE.488529}
}

@article{Ataloglou2023,
  author={Vasileios G. Ataloglou and Sajjad Taravati and George V. Eleftheriades},
  title={Metasurfaces: physics and applications in wireless communications},
  journal={Natl. Sci. Rev.},
  year={2023},
  volume={10}, pages={nwad164},
  doi={10.1093/nsr/nwad164}
}

@article{Zhao2022,
  author={Xiaoguang Zhao and Zhenci Sun and Lingyun Zhang and Zilun Wang and Rongbo Xie and Jiahao Zhao and Rui You and Zheng You},
  title={Review on Metasurfaces: An Alternative Approach to Advanced Devices and Instruments},
  journal={Research},
  year={2022},
  
  doi={10.34133/2022/9765089}
}

@article{Lebbe2022,
  author={Nicolas Lebbe and Kim Pham and Agn{\`e}s Maurel},
  title={Stable {GSTC} Formulation for {Maxwell}'s Equations},
  journal={IEEE Trans. Antennas Propag.},
  year={2022},
  volume={70},
  doi={10.1109/TAP.2022.3161436}
}

@article{HansenMiskovic2026,
  author={Asbjorn Hansen and Zoran L. Mi{\v{s}}kovi{\'c}},
  title={An exploration of the electromagnetic boundary conditions for two-dimensional materials with out-of-plane polarization},
  journal={Appl. Phys. B},
  year={2026},
  volume={132}, pages={33},
  doi={10.1007/s00340-026-08630-3}
}

@article{Phaneuf2025,
  author={Mario Phaneuf and Puyan Mojabi},
  title={Mapping Local Cascaded Metasurface Pairs to Single Nonlocal Metasurfaces},
  journal={IEEE Antennas Wireless Propag. Lett.},
  year={2025},
  volume={24}, pages={354--358},
  doi={10.1109/LAWP.2024.3496432}
}

@article{Phaneuf2026,
  author={Mario Phaneuf and Puyan Mojabi},
  title={Reconciling the Apparent Contradiction of Strongly Dispersive Magneto-Electric Coupling},
  journal={IEEE Trans. Antennas Propag.},
  year={2026},
  
  doi={10.1109/TAP.2026.3746051}
}

@article{Valagiannopoulos2025,
  author={Constantinos Valagiannopoulos},
  title={Causality Implications for Absorption by {EM} Metasurfaces},
  journal={Nanomaterials},
  year={2025},
  volume={15}, pages={793},
  doi={10.3390/nano15110793}
}

@misc{Hansen2026,
  author={Asbjorn Hansen},
  title={Asymptotic electromagnetic transition conditions for thin layers: the arithmetic-mean field and susceptibility hybridization},
  year={2026},
  eprint={2611.04324},
  archivePrefix={arXiv},
  primaryClass={physics.optics},
  note={arXiv:2611.04324 (preprint)}
}

@article{Monticone2025Roadmap,
 author={Monticone, Francesco and Mortensen, N. Asger and others},
 title={Nonlocality in photonic materials and metamaterials: roadmap},
 journal={Opt. Mater. Express},
 volume={15}, pages={1544--1709}, year={2025},
 doi={10.1364/OME.559374}
}

@article{Rahmeier2023,
 author={Rahmeier, Jo{\~a}o Guilherme Nizer and Smy, Tom J. and Dugan, Jordan and Gupta, Shulabh},
 title={Zero Thickness Surface Susceptibilities and Extended {GSTCs}---Part I: Spatially Dispersive Metasurfaces},
 journal={IEEE Trans. Antennas Propag.},
 volume={71}, pages={5909--5919}, year={2023},
 doi={10.1109/TAP.2022.3164169}
}

@article{Ghaneizadeh2024,
 author={Ghaneizadeh, Alireza and Joodaki, Mojtaba},
 title={Generalized Sheet Transition Conditions ({GSTCs}) in Electromagnetic Metasurface Modeling},
 journal={IEEE Access}, volume={12}, pages={74305--74326}, year={2024},
 doi={10.1109/ACCESS.2024.3406160}
}

@article{Achouri2023Multipolar,
 author={Achouri, Karim and Tiukuvaara, Ville and Martin, Olivier J. F.},
 title={Multipolar Modeling of Spatially Dispersive Metasurfaces},
 journal={IEEE Trans. Antennas Propag.}, year={2023},
 doi={10.1109/TAP.2022.3213907}
}

@article{Achouri2023Symmetry,
 author={Achouri, Karim and Tiukuvaara, Ville and Martin, Olivier J. F.},
 title={Spatial symmetries in nonlocal multipolar metasurfaces},
 journal={Adv. Photonics}, volume={5}, pages={046001}, year={2023}
}

@article{Yucel2025,
 author={Y{\"u}cel, Mustafa and Cuesta, Francisco Sebastian and Achouri, Karim},
 title={Angle-invariant scattering in metasurfaces},
 journal={Phys. Rev. B}, volume={111}, pages={235448}, year={2025},
 doi={10.1103/q4wc-64kl}
}

@article{Allahverdizadeh2025,
 author={Allahverdizadeh, Hossein and Achouri, Karim},
 title={Multipolar angular scattering of substrated metasurfaces},
 journal={Phys. Rev. B}, volume={112}, pages={245420}, year={2025},
 doi={10.1103/9cm1-fy6c}
}

@article{Cuesta2025,
 author={Cuesta, Francisco Sebastian and Achouri, Karim},
 title={A General Susceptibility-Based Approach for Uniform Metasurface Scattering},
 journal={Adv. Theory Simul.}, year={2025},
 doi={10.1002/adts.202500698}
}

@article{Welters2014,
 author={Welters, Aaron and Avniel, Yehuda and Johnson, Steven G.},
 title={Speed-of-light limitations in passive linear media},
 journal={Phys. Rev. A}, volume={90}, pages={023847}, year={2014},
 doi={10.1103/PhysRevA.90.023847}
}

@article{Skaar2019,
 author={Skaar, Johannes and H{\aa}genvik, Hans Olaf and Dirdal, Christopher A.},
 title={Four definitions of magnetic permeability for periodic metamaterials},
 journal={Phys. Rev. B}, volume={99}, pages={064407}, year={2019},
 doi={10.1103/PhysRevB.99.064407}
}

@article{Dirdal2015,
 author={Dirdal, Christopher A. and Bondevik, Tarjei and Skaar, Johannes},
 title={Relaxed dispersion constraints and metamaterial effective parameters with physical meaning for all frequencies},
 journal={Phys. Rev. B}, volume={91}, pages={134102}, year={2015},
 doi={10.1103/PhysRevB.91.134102}
}

@article{Bernland2011,
 author={Bernland, Anders and Luger, Annemarie and Gustafsson, Mats},
 title={Sum rules and constraints on passive systems},
 journal={J. Phys. A: Math. Theor.}, volume={44}, pages={145205}, year={2011},
 doi={10.1088/1751-8113/44/14/145205}
}

@incollection{Nedic2021,
 author={Nedic, Mitja and Ehrenborg, Casimir and Ivanenko, Yevhen and Ludvig-Osipov, Andrei and Nordebo, Sven and Luger, Annemarie and Jonsson, Lars and Sj{\"o}berg, Daniel and Gustafsson, Mats},
 title={Herglotz functions and applications in electromagnetics},
 booktitle={Advances in Mathematical Methods for Electromagnetics},
 editor={Kobayashi, K. and Smith, P. D.},
 publisher={Institution of Engineering and Technology}, pages={491--514}, year={2021},
 doi={10.1049/SBEW528E_ch20}
}

@article{BudhuGrbic2021,
  author = {Budhu, Jordan and Grbic, Anthony},
  title = {Recent advances in bianisotropic boundary conditions: theory, capabilities, realizations, and applications},
  journal = {Nanophotonics},
  year = {2021},
  volume = {10},
  number = {16},
  pages = {4075--4112},
  doi = {10.1515/nanoph-2021-0401}
}

@article{Szymanski2021,
  author = {Szymanski, Luke and others},
  title = {Fundamentals of Lossless, Reciprocal Bianisotropic Metasurface Design},
  journal = {Photonics},
  volume = {8},
  number = {6},
  pages = {197},
  year = {2021},
  doi = {10.3390/photonics8060197}
}

@article{Monti2021MultipolarImpedance,
  author = {Monti, Alessio and Al{\`u}, Andrea and Toscano, Alessandro and Bilotti, Filiberto},
  title = {Design of High-Q Passband Filters Implemented Through Multipolar All-Dielectric Metasurfaces},
  journal = {IEEE Trans. Antennas Propag.},
  volume = {69}, number = {8}, pages = {5142--5147}, year = {2021},
  doi = {10.1109/TAP.2020.3045795}
}

@article{Arnold2025,
 author={Arnold, K. O. and Hooper, C. and Smith, J. G. and Hibbins, A. P. and Sambles, J. R. and Horsley, S. A. R. and Clow, N.},
 title={Finding passive, reciprocal metasurfaces for arbitrary wave transformations},
 journal={Phys. Rev. Applied}, volume={24}, pages={044034}, year={2025},
 doi={10.1103/8g1m-hjbz}
}

@article{Srivastava2021,
 author={Srivastava, Ankit},
 title={Causality and passivity: From electromagnetism and network theory to metamaterials},
 journal={Mech. Mater.}, volume={154}, pages={103710}, year={2021},
 doi={10.1016/j.mechmat.2020.103710}
}

@article{Asadchy2016,
 author={Asadchy, V. S. and Albooyeh, M. and Tcvetkova, S. N. and D{\'i}az-Rubio, A. and Ra'di, Y. and Tretyakov, S. A.},
 title={Perfect control of reflection and refraction using spatially dispersive metasurfaces},
 journal={Phys. Rev. B}, volume={94}, pages={075142}, year={2016},
 doi={10.1103/PhysRevB.94.075142}
}

@article{Yaghjian2013,
 author={Yaghjian, A. D. and Al{\`u}, Andrea and Silveirinha, M. G.},
 title={Homogenization of spatially dispersive metamaterial arrays in terms of generalized electric and magnetic polarizations},
 journal={Photonics Nanostruct. Fundam. Appl.}, volume={11}, pages={374--396}, year={2013},
 doi={10.1016/j.photonics.2013.04.001}
}

@article{Kuester2003,
 author={Kuester, E. F. and Mohamed, M. A. and Piket-May, M. and Holloway, C. L.},
 title={Averaged transition conditions for electromagnetic fields at a metafilm},
 journal={IEEE Trans. Antennas Propag.}, volume={51}, pages={2641--2651}, year={2003},
 doi={10.1109/TAP.2003.817560}
}

@article{Albooyeh2016,
 author={Albooyeh, M. and Tretyakov, S. and Simovski, C.},
 title={Electromagnetic characterization of bianisotropic metasurfaces on refractive substrates: General theoretical framework},
 journal={Ann. Phys. (Berlin)}, volume={528}, pages={721--737}, year={2016},
 doi={10.1002/andp.201600015}
}

@article{Albooyeh2017,
 author={Albooyeh, M. and Kwon, D.-H. and Capolino, F. and Tretyakov, S. A.},
 title={Equivalent realizations of reciprocal metasurfaces: Role of tangential and normal polarization},
 journal={Phys. Rev. B}, volume={95}, pages={115435}, year={2017},
 doi={10.1103/PhysRevB.95.115435}
}

@book{Anderson1973,
 author={Anderson, B. D. O. and Vongpanitlerd, S.},
 title={Network Analysis and Synthesis: A Modern Systems Theory Approach},
 publisher={Prentice-Hall}, address={Englewood Cliffs, NJ}, year={1973}
}

@article{Kurokawa1965,
 author={Kurokawa, K.},
 title={Power waves and the scattering matrix},
 journal={IEEE Trans. Microwave Theory Tech.}, volume={13}, pages={194--202}, year={1965},
 doi={10.1109/TMTT.1965.1125964}
}

@article{Sweeney2020,
 author={Sweeney, William R. and Hsu, Chia Wei and Stone, A. Douglas},
 title={Theory of reflectionless scattering modes},
 journal={Phys. Rev. A}, volume={102}, pages={063511}, year={2020},
 doi={10.1103/PhysRevA.102.063511}
}

@book{Harrington2001,
  author = {Harrington, Roger F.},
  title = {Time-Harmonic Electromagnetic Fields},
  publisher = {Wiley-IEEE Press},
  year = {2001},
  note = {Reissue of the 1961 edition}
}

@book{HornJohnson2013,
  author = {Horn, Roger A. and Johnson, Charles R.},
  title = {Matrix Analysis},
  edition = {2},
  publisher = {Cambridge University Press},
  year = {2013}
}

@article{Rahmeier2024TimeDomain,
 author={Rahmeier, Jo{\~a}o G. Nizer and Dugan, Jordan and Smy, Tom J. and Gupta, Shulabh},
 title={Time-Domain Analysis of Temporally and Spatially Dispersive Metasurfaces in {GSTC-FDTD} Frameworks},
 journal={IEEE Access}, volume={12}, pages={22557--22572}, year={2024},
 doi={10.1109/ACCESS.2024.3364518}
}

@article{Dugan2024Synthesis,
 author={Dugan, Jordan and Smy, Tom J. and Monticone, Francesco and Gupta, Shulabh},
 title={Surface Susceptibility Synthesis of Spatially Dispersive Metasurfaces for Space Compression and Spatial Signal Processing},
 journal={IEEE Trans. Antennas Propag.}, volume={72}, pages={6572--6583}, year={2024},
 doi={10.1109/TAP.2024.3426072}
}

@article{Zhuravlev2025Nonlocal,
 author={Zhuravlev, Alexander and Kurenkov, Yury and Wang, Xuchen and Dushko, Fedor and Zalipaev, Viktor and Glybovski, Stanislav},
 title={Radiation-pattern synthesis with uniform nonlocal metasurfaces},
 journal={Phys. Rev. Applied}, volume={23}, pages={044052}, year={2025},
 doi={10.1103/PhysRevApplied.23.044052}
}

@misc{Kouroublakis2026Transient,
 author={Kouroublakis, Minas and Tsitsas, Nikolaos L. and Leviatan, Yehuda},
 title={A Time-Domain Method of Auxiliary Sources for Analyzing Transient Electromagnetic Interactions with {GSTC}-Modeled Metasurfaces},
 year={2026}, eprint={2605.08047}, archivePrefix={arXiv},
 note={arXiv:2605.08047 (preprint)}
}

@article{Khezripour2026Impedance,
 author={Khezripour, Zahra and Abbasi-Arand, Bijan},
 title={Analytical surface-impedance modeling of disordered dielectric metasurface monolayers},
 journal={Sci. Rep.}, volume={16}, pages={25623}, year={2026},
 doi={10.1038/s41598-026-56140-1}
}

@article{Fathnan2020Bandwidth,
 author={Fathnan, Ashif A. and Olk, Andreas E. and Powell, David A.},
 title={Bandwidth limit and synthesis approach for single resonance ultrathin metasurfaces},
 journal={J. Phys. D: Appl. Phys.}, volume={53}, pages={495304}, year={2020},
 doi={10.1088/1361-6463/abb390}
}
\end{document}